\documentclass[nonacm,acmsmall]{acmart}
\usepackage{mathtools}
\usepackage{tikz}
\usepackage{multirow}
\usepackage{balance}
\usepackage{adjustbox}
\usepackage{wrapfig}
\usepackage{tikz}

\usepackage{colortbl}

\usepackage{pifont}
\newcommand{\cmark   }{\ding{51}}%

\newcolumntype{R}[1]{>{\raggedleft\let\newline\\\arraybackslash\hspace{0pt}}m{#1}}
\usepackage{dcolumn}
\newcolumntype{d}[1]{D..{#1}}
\def\sym#1{\ifmmode^{#1}\else\(^{#1}\)\fi}
\newcommand\mc[1]{\multicolumn{1}{c}{#1}}

\newcommand{\parc}[1]{\noindent{\textbf{#1}}\;}

\begin{document}
\title{On the Use of Proxies in Political Ad Targeting}

\author{Piotr Sapiezynski}
\authornote{These authors contributed equally to this research.}
\authornote{Corresponding author.}
\email{p.sapiezynski@northeastern.edu}
\author{Levi Kaplan}
\authornotemark[1]
\email{kaplan.l@northeastern.edu}
\author{Alan Mislove}
\email{amislove@ccs.neu.edu}
\affiliation{%
  \institution{Northeastern University}
  \city{Boston}
  \state{MA}
  \postcode{02115}
  \country{USA}
}
\author{Aleksandra Korolova}
\email{korolova@princeton.edu}
\affiliation{%
    \institution{Princeton University}
    \city{Princeton}
    \state{NJ}
    \postcode{08540}
    \country{USA}
}


\begin{abstract}
  Detailed targeting of advertisements has long been one of the core offerings of online platforms. 
  Unfortunately, malicious advertisers have frequently abused such targeting features, with results that range from violating civil rights laws to driving division, polarization, and even social unrest. 
  Platforms have often attempted to mitigate this behavior by removing targeting attributes deemed problematic, such as inferred political leaning, religion, or ethnicity. 
  In this work, we examine the effectiveness of these mitigations by collecting data from political ads placed on Facebook in the lead up to the 2022 U.S. midterm elections. 
  We show that major political advertisers circumvented these mitigations by targeting proxy attributes: seemingly innocuous targeting criteria that closely correspond to political and racial divides in American society.
  We introduce novel methods for directly measuring the skew of various targeting criteria to quantify their effectiveness as proxies, and then examine the scale at which those attributes are used.
  Our findings have crucial implications for the ongoing discussion on the regulation of political advertising and emphasize the urgency for increased transparency.  
\end{abstract}





\received{July 2023}
\received[revised]{April 2024}
\received[accepted]{July 2024}
\maketitle

\section{Introduction}
Online social media platforms, including Facebook, Google and TikTok, have developed powerful advertising services based on \textit{targeted advertising}, by which an advertiser can specify which segment of the user base they want to reach~\cite{Facebook_target_2023}.
These targeting criteria are significantly more fine-grained than what was possible through previous forms of advertising (e.g., newspapers or television).  These targeting criteria often includes targeting options ranging from demographic characteristics, to location, education, ethnic affinity, political affiliation, and many thousands of interests, both self-declared by the users and inferred by the platform \cite{speicher2018potential}.
The resulting capability to select narrow groups of users is welcome by the advertisers but also raises societal concerns in domains ranging from life opportunity advertising (e.g.,~housing, credit, and jobs)~\cite{Angwin_Parrisjr_2016, Angwin_Scheiber_Tobin_2017,sapiezynski2022algorithms} to health~\cite{Lecher_2021}.

In this paper, we focus on targeted {\em political} advertising on Facebook, which has raised similar concerns~\cite{ribeiro2019microtargeting, zuiderveen2018online, mueller2019report, kim2018stealth, barocas2012price, tufekci2014engineering, ali2021ad}. Fine-grained targeting has been shown to enable voter manipulation and misinformation, by allowing political actors to craft different and manipulative ad messages for specific audiences on Facebook, unbeknownst to the recipients and without oversight~\cite{The_Guardian_2018, mueller2019report}. 
Despite these controversies, Facebook has remained steadfast in their commitment to allowing political advertisements on the platform; arguing that they are essential to a free and open democracy~\cite{Clegg_2019}.
Although Facebook blocked the creation of new political ads in the U.S.~the week before the 2022 U.S.~midterm elections, it allowed preexisting ads to continue running and quickly rescinded the ban on new ads after the election.
Facebook's primary response to concerns regarding the abuse of fine-grained targeting in political advertising took the form of creating the Facebook Ad Library \cite{Constine_2019} and removal of a number of targeting capabilities~\cite{MetaBusiness_2021}.
Targeting criteria removed to date include the ability to target by ethnicity\footnote{Facebook had previously allowed targeting by ``ethnic affinity'' rather than ethnicity explicitly. The categories included ``African-American Affinity'', ``Hispanic Affinity'', and ``Asian Affinity''. ``White'' was not an available category.} and political affiliation, as well as a number of interests and topics that Facebook deems sensitive, such as health causes, sexual orientation, religious practices and groups, and social issues, causes, organizations, and public figures~\cite{MetaBusiness_2021}.
Importantly, Facebook did not provide a comprehensive list of what targeting options they removed, and left the possibility for Facebook to continue to review, update and remove targeting options.

In this work, we investigate whether the approach taken by Facebook---that of removal of particular targeting capabilities---achieves its intended public interest goals, or whether advertisers are able to work around this restriction.
Specifically, it has long been known---including by Facebook itself \cite{fb_data_science}---that certain facially-neutral interests can serve as \textit{proxies} for political affiliation, race, or gender.
For example, most users with an interest in Country music are white, whereas most users with an interest in R'n'B music are Black.
Thus, even though musical preferences are facially neutral, when used for targeting ads they can serve as efficient proxies for race, and thus, enable exclusion of certain racial groups from particular kinds of advertising. 

We focus on the case of political advertising, and study whether, in response to removal of explicit political interest targeting categories by Facebook, the political advertisers have moved to targeting by proxies instead.
We do so in three steps.
First, we develop methodologies to measure which interests serve as effective proxies for political affiliation and ethnic affinity without access to any proprietary Facebook data, and without relying on the names of interests.
Second, we crawl the Facebook Ad Library to collect targeting information from all advertising accounts which ran political ads during the run-up to the U.S.~2022 midterm election.
We focus on political advertising because that is the only category of advertisers for whom Facebook makes {\em targeting} data available in its Ad Library.
Finally, we utilize the methodologies we develop and data we obtain to measure the extent to which political advertisers rely on proxy attributes to circumvent the targeting limitations introduced by Facebook during the U.S.~2022 midterm elections.
We hypothesize that the trends for proxy use apply also to advertisers in other categories, such as employment, housing, and health.

\subsection{Contributions} 
Our contributions can be summarized as follows:

\parc{Novel methodologies for measuring skews of targeting parameters to establish their proxy power.}
In the five weeks leading up to the 2022 midterm elections, political advertisers used over 19,000 unique interests, demographics, and behaviors to target ads.
A core challenge in identifying which of these were effectively proxies for the targeting options removed by Facebook is the measurement of the political and racial \textit{skew} of these varied targeting criteria at scale, without reliance on proprietary datasets, and without reliance on the names of these criteria.
To address this challenge, we develop and implement two novel measurement approaches.
They both rely on the use of publicly available data combined with non-traditional use of Facebook advertiser tools, and they do not require any privileged access to internal Facebook data.
Moreover, they do not require running ads or paying Facebook, making the identification of highly skewed interests at scale feasible to other public interest researchers.

The first approach allows for a direct measurement of popularity of different targeting criteria among politically and racially uniform Custom Audiences.
The second approach combines data obtained through the Facebook Audience Insights tool---which reports Facebook pages popular among users associated with each interest---with an external dataset containing measures of political biases of different Internet domains.
Crucially, neither of these approaches relies on interpreting the labels that the platform uses to describe the targeting interests. 
Instead, we report qualitative measures of the efficacy of each interest as a proxy for a variable of interest.
The skew measurement results of both approaches are highly correlated, demonstrating their viability and providing confirmation that they are indeed measuring the potential for these targeting features to serve as proxies.

\parc{Documenting the use of proxies in the wild.}
We then examine the extent to which real political advertisers target and exclude particular sub-populations via proxies. 
We find over 22,000 unique advertising accounts with active political ads in the five weeks leading up to the 2022 midterm elections.
We show advertisers spent millions of dollars targeting or excluding audiences using interests with strong partisan and/or race skews.
Such behavior is problematic both because it is accomplishing things that are, under the current ad targeting system design, intended to be forbidden and because, by doing so via proxies, it makes the behavior harder to detect.

\smallskip
Taken together, our findings demonstrate that the current approach taken by platforms (and Facebook in particular) to prevent targeting by race or political leaning through the removal of select targeting criteria is ineffective, and that advertises can---and do---circumvent the protections put in place with some sophistication and negligible cost.
Using political advertising as a case study, we demonstrate that proxy targeting is widely used by actors on both sides of the U.S.~political spectrum.
Our findings, and the process of obtaining them, motivate the need for increased transparency from the ad platforms as well as regulation that would require a radically different approach to preventing abuse of fine-grained ad targeting capabilities in domains of societal importance.

\section{Background and Related Work}
The fine-grained targeting options offered by Facebook enable both benign and malicious advertisers to reach individuals with unparalleled precision.
In this section, we first provide the necessary background on targeting capabilities offered by Facebook to advertisers.
We detail notable examples of use and abuse of the targeting features that led to spreading of hate speech, discrimination in access to opportunities, or political manipulation.
We highlight the steps that legislators and Facebook have taken to address those harms, such as limiting the targeting capability and progressive introducing of transparency mechanisms.

\subsection{Online targeting capabilities}
Traditional print media and television advertising allows for rudimentary targeting of broad audiences by placing ads strategically in outlets particularly popular among the target audience.
For example, an advertiser seeking attention from men could place an ad in the \textit{Men's Health} magazine and reach an audience of predominantly men. 
Online advertising, on the other hand, enables far more fine-grained targeting of individuals. 
Advertisers can narrow down the target audience by describing their location (from countries down to particular neighborhoods), demographic attributes (including age, gender, education, or marital and parental status), as well as both self-reported, and inferred interests (more than 50,000 unique interests were available for targeting on Facebook at the time of this work)~\cite{Facebook_target_2023}.
To further increase the breadth of targeting criteria, Facebook has also purchased inferences about their users from third parties in the past~\cite{venkatadri2019auditing}.
Moreover, advertisers can use a tool called \textit{Custom Audiences} to target particular individuals directly through personally identifiable information (PII), such as names and addresses, device/advertising IDs, emails, or phone numbers~\cite{Facebook_custom_audiences_2023}.
When using Custom Audiences, the advertiser uploads a list containing PII of individuals; Facebook matches this information with actual Facebook users and delivers the ads to a subset of them.
A \textit{Lookalike Audience} feature allows the advertiser to direct Facebook to create an audience of 
users \textit{similar} to the users in the Custom Audience.
The definition of that ``similarity'' is not public, but previous work showed that Lookalike Audiences carry over demographic and political biases from the custom audiences they were built upon~\cite{speicher2018potential,sapiezynski2022algorithms}.
Combining various forms of targeting through boolean logic further extends the advertisers' ability to precisely target users~\cite{use_detailed_targeting}. 
For example, advertisers can narrow down a Custom Audience to users who have a specific interest and fall into a particular age group.
Furthermore, all of these tools can also be used for \textit{excluding}, rather than targeting, i.e. specifying the individuals who should \textit{not} be shown the ad. 

Despite the vast data collection and inference apparatus, multiple studies reported that users find a significant fraction of inferences made about them inaccurate~\cite{venkatadri2019auditing,sabir2022analyzing}.
Nevertheless, the accuracy of these inferences is irrelevant for their effectiveness as tools for micro-targeting, as long as they enable advertisers to reach homogeneous groups. 
For example, if Facebook infers that Black users are much more likely than white users to be interested in ``The Breakfast Club (movie)'' that interest can be used as an effective proxy for race, despite the fact that this inference is likely wrong.\footnote{This example is not hypothetical; we found that the interest ``The Breakfast Club (movie)'' is indeed predominantly associated with Black users on Facebook.  We note that the The Breakfast Club is also a name of a radio show that is unrelated to the 1985 movie The Breakfast Club (to which the name of the targeting interest refers).  We hypothesize that Facebook's inference methodology is incorrectly associating interest in the radio show with the movie, but regardless, the attribute serves as an effective proxy.}

\subsection{Ad delivery optimization}\label{sec:ad_delivery} The process of audience selection does not end with the advertiser's targeting decisions. 
Because ad budgets are generally not high enough for the ad to reach every user in the targeted audience, Facebook strategically sub-selects who among the targeted audience will actually see the ad.
This choice is not random; instead, it is made in a way that optimizes both the stated goal of the advertiser (for example, maximizing the number of clicks, or product purchases, or impressions received), and Facebook's optimization objectives, such maximizing user time spent on the platform.
This process, referred to as ``ad delivery optimization'', has unfortunately been shown to produce discriminatory effects through skewed distribution of opportunity ads~\cite{ali2019discrimination,imana2021auditing,kaplan2022measurement, imana2024auditing}.
In the case of political ads, \citet{ali2021ad} have shown that the ad delivery optimization can lead to price discrimination based on whether a politician's political leaning is deemed aligned with that of the target audience by Facebook's machine learning algorithms.
This means that the delivery algorithm may charge two political advertisers vastly different amounts to show their ads to the same target audience at the same time, effectively limiting the diversity of political messages shown to the users.

\subsection{Micro-targeting}
The practice of carefully targeting political ads predates social media.
Even prior to the arrival of Facebook and the proliferation of data brokers, political strategists would purchase customer information from loyalty programs, subscriptions, online stores, or car dealers~\cite{edsall2006democrats,hamburger2005parties,murray2010microtargeting}.
In combination with voter records, these datasets enabled them to identify proxy attributes, such as which car brands, or alcohol types were preferred by voters in each party.
For example, in 2004, the successful reelection of George W.~Bush was at least partially attributed to the outreach targeting individuals identified using such methods~\cite{hamburger2005parties}.
The 2008 presidential campaign of Barrack Obama relied heavily on combining the above mentioned data sources with extensive surveys, resulting in rich profiles of individual voters. 
These profiles were then used to predict susceptibility of individuals to different conversation scripts and forms of outreach~\cite{mit2012obama}.
Despite the rapidly growing emphasis on data-driven voter mobilization, online advertising remained a minuscule fraction of political advertising budgets in 2012, below 1.7\%~\cite{cassino2017final}.

However, advertising platforms began collecting significant amounts of data and introducing powerful targeting features to advertisers.
This had the effect of democratizing access to targeted advertising:
rather than each political campaign having to obtain and develop its own data to identify particular kinds of constituents, their propensity to vote, susceptibility to particular messages, Facebook performed all the data mining, inferences, and packaging, and offered it as a service. Moreover, Facebook did so based on all profile and activity data available about each individual, on and off the platform, rather than being limited to the more coarse grained information in voter records and loyalty programs.  Facebook even provided dedicated employees to help political campaigns take advantage of its tools~\cite{guardian2017brad}. 

The use of micro-targeting for political advertising entered the public consciousness during the 2016 U.S.~presidential election. 
First, the winning campaign of Donald J.~Trump relied heavily on targeted advertising on Facebook.
The campaign ran up to 175,000 variants of the same ad per day~\cite{lapowsky2018facebook}, and did not deploy traditional TV advertising until late in the campaign~\cite{miller2016tvads}. 
Importantly, the campaign used the available tools to reach Democratic voters in order to dissuade them from voting altogether~\cite{green2016inside}.
The digital director of Trump's campaign, Brad Parscale, went as far as to say that ``Facebook was going to be how he [Trump] won'', and praised the ability to target small, specific audiences, otherwise unreachable using traditional advertising~\cite{guardian2017brad}.
Second, the Russian company Internet Research Agency (IRA) ran thousands of targeted Facebook ads in an attempt to sow division in the American public, lower African American turnout, and thus bolster Trump's campaign~\cite{mueller2019report}.
IRA relied on micro-targeting via proxy criteria whose names indicate the demographics they tried to reach: ``African-American history'', ``African-American Civil Rights Movement'', ``Malcolm X'', etc. to target Black Americans, and ``Chicano Movement'', ``Chicano rap'', ``Hispanidad'', etc. to reach Hispanic/Latinx users~\cite{ribeiro2019microtargeting}.
Third, two years after the election, a whistleblower revealed that Trump's campaign also retained services of Cambridge Analytica, which collected large amounts of data on Facebook users and then constructed ``psychological profiles'' of these unsuspecting users and craft ads that would exploit their particular susceptibility to persuasion~\cite{matz2017psychological}.

As a result, lawmakers have attempted to take steps to constrain such political micro-targeting.  
For example, the proposed Honest Ads Act in the U.S.~would have forced platforms to publish targeting information for all ads in their libraries~\cite{Honest_Ads_Act}, but it was never passed. 
A recently-approved E.U.~regulation on political advertising will require platforms to seek explicit user opt-in for processing their data for targeting of political ads~\cite{lomas2023ads}.
Data that can reveal a users revealing a racial/ethnic origin, political opinion, religious beliefs, or sexual orientation, will not be permitted for targeting and delivery optimization (see Section~\ref{sec:ad_delivery}). 
However, the law stops shy of completely banning microtargeting. 

\subsection{Platform policies}

After the 2016 U.S.~presidential election, much research explored risks that micro-targeted political advertising poses to democracy and human rights.  These include the risks of sealing people into echo-chambers while limiting their access to pluralistic information; facilitating campaigns' abilities to misrepresent their positions by sending tailored messages with different focus to narrow groups; and enabling the spread of mis- and dis-information, the discouragement of civil participation, and the manipulation of voters~\cite{CLUFE2021, fbemployees2019, EUwhy2023, borgesius2018online}. 
As a result, in the years that followed, many social media platforms changed their policies to attempt to address the risks of micro-targeted political advertising. 
For example, TikTok explicitly banned all political and issue ads~\cite{Chandlee_2019}; Twitter banned political ads;\footnote{Although, following change in leadership, it announced in January 2023 that it would reinstate ``cause-based ads''~\cite{Conger_2023}.} and Google only allowed political ad targeting by location, age, and gender, with interest or custom audience targeting prohibited~\cite{Google_2023}.

Facebook, however, has not banned political advertising or selected only a few permissible targeting criteria.
Instead, Facebook's changes to its advertising system have been largely driven by legal challenges that allege violations of U.S.~anti-discrimination law.
We systematize the detailed timeline of these changes in  Figure~\ref{tab:timeline}, and note a few of the key events here.

\begin{table}
    \small
    \centering
    \renewcommand{\arraystretch}{1.2}
    \begin{tabular}{rp{12cm}}
    \textbf{Time} & \textbf{Action undertaken by Facebook} \\ \hline
    {Feb 2017} & Set to disapprove opportunity ads that use multicultural affinity segments~\cite{Facebook_target_2017}. \\
    {Nov 2017} & Undertook review of exclusion, focusing on the use of exclusion for multicultural affinity groups and other
    potentially sensitive segments (e.g., segments that relate to the LGBT
    community or to religious groups)~\cite{Facebook_target_2017b}. \\
    {Apr 2018} & Removed thousands of categories from exclusion targeting; focused mainly on topics that relate to potentially sensitive personal attributes, such as race, ethnicity, sexual orientation and religion~\cite{Facebook_target_2018}. \\
    {May 2018} & Introduced the Ad Library to archive all political ads~\cite{Leathern_2018}. \\
    {Aug 2018} & Removed 5,000 targeting options to help prevent misuse. Includes limiting the ability for advertisers to exclude audiences that relate to attributes such as ethnicity or religion~\cite{Facebook_target_2018b}. \\
    {Mar 2019} & A settlement with civil rights organizations requires removing gender, age, and ethnic affinity from opportunity ads targeting, along with features that describe or appear to be related to race, color, national origin, ethnicity, gender, age, religion, family status, disability, and sexual orientation~\cite{aclu_2019}. \\
    {Aug 2020} & Removed some targeting options such as multicultural affinity segments~\cite{Facebook_target_2020}. \\
    {Jan 2022} & Removed detailed targeting options that relate to topics people may perceive as sensitive, such as options referencing causes, organizations, or public figures that relate to health, race or ethnicity, political affiliation, religion, or sexual orientation~\cite{Facebook_target_2022, Facebook_target_2022b}. \\
    {May 2022} & Introduced aggregated targeting information to the Ad Library~\cite{king2022targeting}. \\
    {Mar 2024} & Removed targeting options that relate to topics people perceive as sensitive (e.g., targeting options referencing causes related to health, race or ethnicity), or because of legal or regulatory requirements~\cite{Facebook_target_2024}.

    \end{tabular}
    
    \caption{Timeline of changes that Facebook introduced to limit micro-targeting.}
    \label{tab:timeline}
\end{table}

After 2016 reporting by ProPublica showed that Facebook's targeting system enabled advertisers to exclude people from housing ads based on their ethnicity~\cite{Angwin_Jr._2016}, Facebook first limited the ability of advertisers to use multicultural affinity targeting as an exclusion criteria~\cite{Egan_2016,  Facebook_target_2017, Facebook_target_2017b, Facebook_target_2018}.
After further pressure from the press~\cite{Merrill_2020, Kofman_Tobin_2019} and a lawsuit from civil rights organizations~\cite{aclu_2019}, Facebook removed more than 5,000 targeting options that ``describe or appear to be related to'' a range of protected characteristics or classes such as ethnicity and religion ~\cite{Facebook_target_2018b, Facebook_target_2020}.
At the same time, civil rights organizations, regulators, and lawmakers pressured Facebook to address the micro-targeting risks of political advertising~\cite{wapost2019, gates2019, uk2018, nyt2021, weintraub2019microtargeting,eshoo2020, eshoo2021, HRW2022, cdt2023open, allen2024eu}.
In 2022, Facebook made the largest change to its available targeting options, affecting all ads. Namely, it removed targeting options ``that relate to topics people may perceive as sensitive, such as options referencing causes, organizations, or public figures that relate to health, race or ethnicity, political affiliation, religion, or sexual orientation'' with examples including options that ``relate to political beliefs, social issues, causes, organizations and figures''~\cite{Facebook_target_2022, Facebook_target_2022b, wsj2021}.

As Facebook continues to remove targeting options---including the most recent update in March 2024~\cite{Facebook_target_2024}---anecdotal reports have pointed to advertisers circumventing these limitations by using proxies~\cite{Keegan_2021,corasaniti2021political,tolan2022political,markay2022americas,whotargetsme2022sweden}.
Our study, therefore, aims to develop a scalable methodology to better understand whether and how political advertisers use proxies, focusing on the U.S. 2022 midterm elections.
Our goal is to determine
in an independent and systematic fashion whether the safeguards Facebook put in place are achieving the desirable societal and democratic goals, and whether accompanying transparency mechanisms implemented by Facebook facilitate such research.

\subsection{Harms beyond political polarization}
We already described how malicious actors can use detailed targeting to sow political division.
Unfortunately, the range of harms that can occur as a result of targeted advertising does not end there.
Here, we describe related documentation of other harms.

\parc{Proliferation of hate speech.} In 2017 ProPublica reported that, prior to their alerting Facebook, the platform enabled targeting several anti-Semitic interest topics such as `Jew hater', `How to burn jews', and `History of `why jews ruin the world' ' (all spelling original) \cite{Angwin_Varner_Tobin_2017}. Presumably, these targeting criteria allowed advertisers to speak directly to extreme anti-Semites. Importantly, these ad categories were algorithmically created, and made available to advertisers without manual inspection.

\parc{Exploiting vulnerability.}
Targeted advertising is perceived by many as manipulative \cite{wu2023slow}, and has been described as a form of ``slow violence'', where the gradual accumulation of exposure to harmful targeted ads can manifest itself as real emotional harm \cite{gak2022distressing}. 
The slow violence of targeted advertising is expanded upon by Wu and colleagues, who describe it as consisting of four harms: psychological distress, loss of autonomy, constriction of user behavior, and algorithmic marginalization and traumatization~\cite{wu2023slow}. 

Platforms enable this harm by exposing sensitive characteristics of users through targeting criteria.
Advertisers can narrow down the users to those who suffer from body-image insecurities, as well as health-related anxieties, and even specific diseases.
Already in 2017, leaked documents obtained by \textit{The Australian} indicated that Facebook was identifying teens feeling ``insecure'', ``worthless'', ``stressed'', ``anxious'' for targeting use \cite{Machkovech_2017}. 
More recently, The Markup found that pharmaceutical companies would target users ``interested'' in awareness of diseases as a proxy for individuals who have that disease \cite{Lecher_2021}. 
For example, cancer treatments were shown to be targeted to users interested in `Cancer Awareness' or `National Breast Cancer Awareness Month' and ads for drugs treating bipolar were targeted at those interested in the `Depression and Bipolar Support Alliance' and the `National Alliance on Mental Illness.' 
Evidence suggests that over-exposure to such triggering content can worsen anxiety symptoms in the vulnerable individuals~\cite{gak2022distressing,panoptykon2021algorithms}, in addition to raising privacy fears from just the fact that Facebook infers such interests. 

\subsection{Efficacy of targeted advertising}
In the previous section we described that micro-targeting continues to be wide-spread and employed by various parties, often with multi-million dollar budgets.
However, systematically demonstrating that narrowly targeted ads actually do work for influencing recipients is difficult.
Observational studies cannot typically be used to identify causality, so special care must be taken in the study design to minimize the effect of confounding variables.
Here, we mention a few studies that did so and showed increased efficacy of targeted ads. 
\citet{matz2017psychological} were the first to perform an experimental validation of targeting people with particular psychological profiles via proxies, i.e., the approach the Cambridge Analytica was later found to be using.
They created ads that would appeal to those with certain psychological traits (for example low extraversion, or high openness,\footnote{The authors relied on a five trait personally model referred to as The Big Five or OCEAN for openness to experience, conscientiousness, agreeableness, extraversion, neuroticism. They had previously shown that these traits can be reliably inferred using Facebook likes~\cite{youyou2015computer}} etc.) and then used proxies to target these individuals.
The results of the experiment indicated users were more likely to click through the ads whose content was congruent with their personally targeted via proxies. 
~\citet{tappin2023quantifying} ran issue ads targeted using demographics and psychological traits, and tested to resonate with the target group.
In their experiment, messages crafted this way performed up to 70\% better than the best performing non-personalized messages. 
While the two studies described above were experimental, \citet{ribeiro2019microtargeting} took a different approach.
They documented IRA's actual use of micro-targeting using proxy attributes in real ads. 
Then, they performed a representative survey and showed that voters whose ethnicity matched that implied by the proxy attributes used for targeting were less likely to identify any misinformation within, and less likely to report these ads. Instead, they were more likely to approve of their content.

\subsection{Ad Library}
\label{sec:ad_library}
In May 2018, Facebook introduced the Ad Library~\cite{ad_library}, an online searchable archive of all political and issue ads that are currently running or have run on the platform since its launch~\cite{Constine_2019}.
The library shows the content of each political ad, the details of the account that ran the ad as well as information on who paid for it, and a range of meta-information: the amount spent and the count of unique impressions, as well as geographic and demographic breakdowns of the audience who were delivered the ad~\cite{Leathern_2018}.
This content is available through the Library's website and a subset of it can also be retrieved through an API, subject to strict rate limits that prevent cataloging of all ads by third parties~\cite{le2022audit}.
As of 2022, the Library also provides aggregated information on targeting parameters used by every political advertiser, but does not make this information available through the official API.  
The reported targeting criteria include: age and gender (\textit{self-reported} by the user), location, whether the advertiser used the Custom Audience~\cite{fb_custom_audience} or Lookalike Audience~\cite{fb_lookalike_audience} feature, and detailed targeting (which includes both \textit{self-reported} and \textit{platform-inferred} interests, demographics, and behaviors). 
Additionally, each targeting criterion is accompanied with the count of ads it was used in, as well as the fraction of the total budget that was spent on ads targeted with that criterion.
Ads that are not classified as political or issue ads are only visible in the Ad Library while they are running, and are not accompanied with the budget, targeting, and delivery information, unlike political ads. 

\subsubsection{Limitations of the Ad Library}
While Facebook did take into account some of the desiderata expressed by legislators in Honest Ads Act~\cite{Honest_Ads_Act} and civil society advocates~\cite{Rieke_Bogen_2018, Madrigal_2018} when designing the Ad Library, its deployment has not been without criticism. 
For example, because of the faulty detection mechanism of which ads are political, the library fails to report significant portions of political advertising on the platform~\cite{edelson2020security, le2022audit}. 
Furthermore, at the time of writing this work, the targeting information is presented in a way that prevents researchers from asking a number of fundamental questions. 
Importantly, the targeting parameters are not grouped per ad, but instead aggregated per week of an advertiser's activity, regardless of how many ads they ran in this period. 
For example, the Ad Library might report that the advertiser ran two ads; and that among these two, one was targeted to users interested in `NPR', and one to users interested in `Duck Dynasty'. 
From the presented information, the researcher does not know whether each of these interest was used to target a different ad (indicating that the advertiser is trying to reach different audiences with different ads), or they were both used to target the same ad (targeting a politically-broad audience with a uniform message).
Finally, there is currently no official API access to the targeting information and researchers have to instead rely on an undocumented API that powers the website, with stringent rate limits.

\subsection{Prior work on measuring skews, alignments, biases and slants in social media}
\label{sec:related-skews}
So far we have referred to various targeting criteria as \textit{proxies} for other characteristics of interest. 
We do not anticipate finding perfect proxies, i.e. interests targeting which would yield an audience where all users share the desired characteristic.
Therefore, we need a quality (purity) measure to quantify how close to perfect the proxies are.
We now provide an overview of related measures introduced in previous work. 

\citet{budak2016fair} set out to quantify \textit{ideological slant} of news stories.
They used crowd-workers to judge, on a five-point scale, the extent to which each story was favorable towards the Democratic party, and then, separately, to which extent it was favorable towards the Republican party.
They defined ideological slant as the difference between the two, scaled from -1 (only favorable to Democrats) to 1 (only favorable to Republicans).

\citet{bakshy2015exposure}, as employees of Facebook and using Facebook's internal data, set out to measure the diversity of view points that Facebook users were exposed to through their algorithmically curated News Feeds.
To that end, they identified news stories which were shared by at least 20 users who self-reported their political alignment in their profiles.
They then averaged these affiliations to create a score ranging from -1 for ``only shared by Democrats'' through 0 for ``shared equally by Democrats and Republicans'' to 1 for ``shared only by Republicans''.
Note that this score does not describe the slant of the news story--instead, it quantifies the alignment of the audience that shared it.

Following a similar methodology, \citet{robertson2018bias} created a panel of Twitter users whose political alignment they learned by matching them to publicly available voter records with self-reported party affiliation.
They identified the domains that the panel of users mentioned in their tweets, and used these users' known affiliation to compute the audience \textit{bias} on a scale from -1 to 1.
Notably, Robertson's et al. metric allows for measuring biases in situations where the two groups are not of similar sizes. 
A zero in their metric indicates that a members of one group are as likely as those in the other group to share a piece of content, not that an even number of members in those groups did so.

Finally, \citet{ribeiro2018media} introduced a method for measuring news outlets' readership bias. 
They were able to use the Facebook API to directly query the number of users of different ethnicities and political views ``interested'' in various media outlets and measure the bias directly. 
Since the removal of targeting by ethnicity and political views, that method can no longer be replicated directly, but it does serve as a starting point to the methods we propose in this paper.

All the methodologies rely on either extensive manual effort to rate partisanship of sources and craft audiences, or access to private datasets, thus making them difficult to maintain, scale, and reproduce.
In contrast, our first approach relies only on data that is publicly and freely available through the Facebook advertising interface and official election websites.

The works described in this section have used different names for related phenomena: slant, alignment, and bias.
Given the nature of our data, none of these metrics translates directly to the problem we set out to measure.
Therefore, throughout this work we will use the term \textit{skew} to emphasize that the formulation is different than the previously described measures.

\section{Data and Methods}
Among the core contributions of this work are two methods of measuring the skews of a targeting interest. 
In this section, we first provide a brief overview of both approaches. 
We then elaborate on techniques for obtaining the necessary data and performing the calculations.
Finally, we reveal the distributions of the skews among interests that were actually used by advertisers during the 2022 U.S. Midterm Elections.

\begin{figure*}[t!]
    \includegraphics[width=1\linewidth]{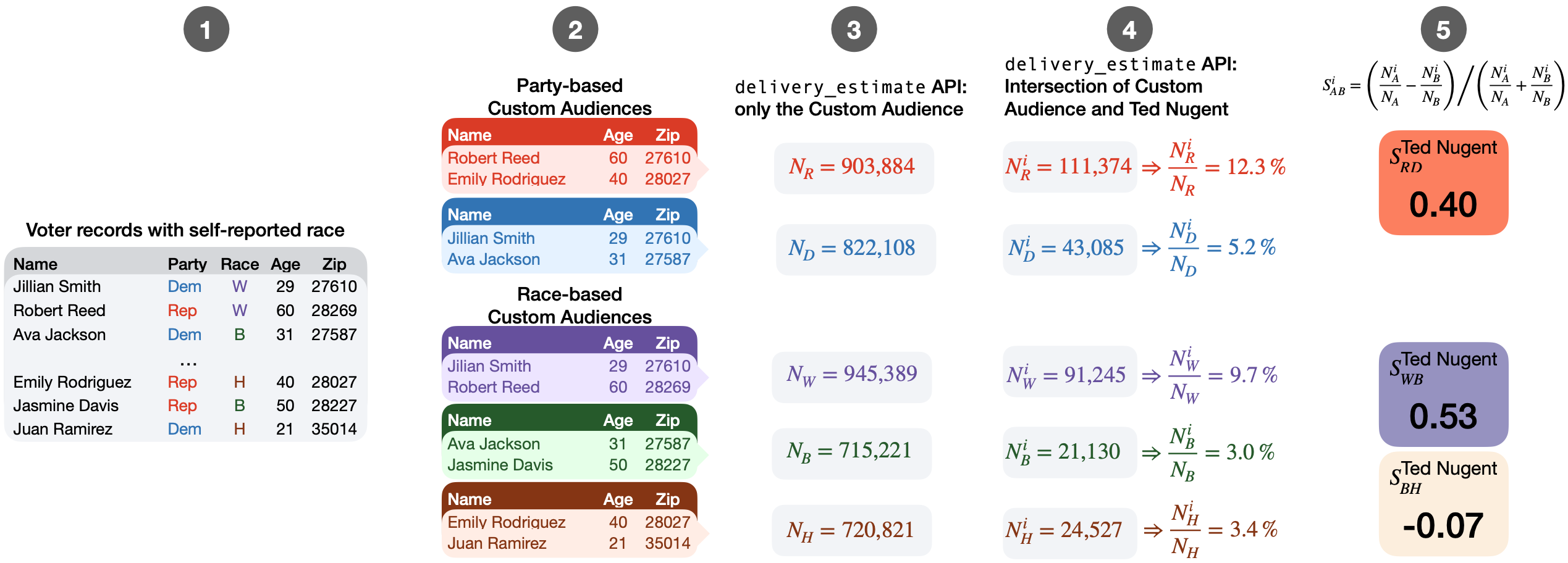}
    \caption{Estimating the skew of targeting interests using the overlap between custom audiences and inferred interests. Step 1: Consolidate voter records from all states where voters self-report race.  Step 2: Create Custom Audience out of party- or race-uniform samples of voters. Step 3: Use the \texttt{delivery\_estimate} API to estimate how many users can actually be reached in each custom audience. Step 4: Use the \texttt{delivery\_estimate} API to estimate what fraction of users in each custom audience share a given interest (here: Ted Nugent). Step 5: Use Eq.~\ref{eq:skew_def} to calculate skew.}
    \label{fig:customs}
\end{figure*}

The \textbf{first approach}, summarized in Figure~\ref{fig:customs}, exploits the Facebook advertising interface. We create politically or demographically homogeneous Custom Audiences
and use the \texttt{delivery\_estimate} API to estimate the number of active users in each. We then use the same API to estimate the fraction of users in each audience that share the particular interest. We compare these fractions using a skew measure we introduce in Section~\ref{sec:data-voter+audience} to assign a numerical value to the interest's skew.
The \textbf{second approach}, depicted in Figure~\ref{fig:insights}, leverages the fact that the Facebook Audience Insights tool reports Facebook Business pages popular among users who share each interest.
Each of these pages represents an entity (a public person, a business, a media outlet, etc.), and reveals their non-Facebook website URL.
These URLs, in turn, can be looked up in the audience skew database published by \citet{robertson2018bias}.
Once we obtain the audience skews \emph{for each of the URLs} associated with a particular interest, we average them to estimate the audience skew \emph{of that interest}.
Both approaches require an interest as input to calculate its skew, so before we describe these approaches in detail, we explain how we obtained a list of interests used by political advertisers.

\subsection{Obtaining targeting interests from the Facebook Ad Library}
\label{sec:targeting-data}
We collected the targeting information from the Ad Library daily in two steps. 
First, we scraped the Meta Ad Library Report which lists all advertisers active during the previous day.\footnote{See: \url{https://www.facebook.com/ads/library/report/}}
Second, using this list, we requested, one by one, the targeting report for each of these advertisers using an undocumented API~\cite{inspect2023browser}.
This work analyses the data obtained during a month leading up to the U.S~midterm elections of 2022, i.e., October 8, 2022 through November 8, 2022.
The data contains information about all 22,479 advertising accounts that ran any political ads to U.S. audiences in that period, even if the ads did not pertain to the midterm elections.
These accounts targeted the audiences using a total of 14,662 unique interests, 4,236 demographic criteria, and 269 behaviors.
Further, the advertisers ran ads whose targeting excluded a total of 1,025 interests, 151 demographic criteria, and 40 behaviors. 
Through the remainder of this paper we will refer to this collection as the \textbf{Political Ad Targeting Dataset}.

Although the targeting information was available when manually loading the advertiser page in the Ad Library, the official API for the same data did not report this info at the time of our data collection. We thus 
relied on an undocumented API, which we identified by analyzing the requests our Internet browser made while browsing the Ad Library~\cite{inspect2023browser}.
We note that Facebook's underlying system and, as a consequence, the undocumented API did not always work in a predicable fashion. 
The delay with which it reported the data varied from two to six days, with 95\% of successful requests returning information dated three or four days before. 
Furthermore, the API did not report any targeting information at all in 4.9\% of our requests, despite the fact that we knew that corresponding ad accounts were indeed running political ads.
We found the vast majority of missing data comes from the same set of accounts, with less than 10\% of successful requests for 3.9\% of accounts.

\subsection{Measuring political and racial skews using voter records}
\label{sec:data-voter+audience}
Our \textbf{first approach} leverages our ability to create racially and politically homogeneous Custom Audiences to then measure the popularity of different targeting criteria among these groups. 
In particular, we obtain voter records\footnote{The data was purchased by our Institution from L2, see: https://l2-data.com/datamapping/. The purchase of The L2 List was made unrelated to this work, but the data is available upon request to all faculty, students, and staff at the institution. 
This research work is a derivative from the use of The L2 List and, per the terms of use, is a property of the authors.
L2 has had no influence on the research direction, nor did they have access to the manuscript before its submission. 
Before transitioning to the L2 dataset, we had used the publicly available voter records for North Carolina and Florida. 
The results we present here can be closely replicated using that data.} 
from the states where voters self-report their race (see Panel~1 of Figure~\ref{fig:customs}): Alabama, Florida, Georgia, Louisiana, North Carolina, South Carolina, and Tennessee. Note that all of these states are in the South, thus the ethnic skews we observe will be more representative of the residents of that region of the US. 
We split these voters into groups such that each one contains only Democratic voters, and another contains only Republican voters; we additionally create one for each ethnic group: white, Black, and Hispanic.
These groups are of different sizes so to avoid biasing the results by discrepancies in coverage. We randomly sample two million individuals from each and upload them as separate Custom Audiences to Facebook, as shown in Panel~2 of Figure~\ref{fig:customs}.

Once we have crafted our Custom Audiences, we turn to another Facebook tool called the Ad Manager interface, and the \texttt{delivery\_estimate} API that underlies it. 
The API provides advertisers with rough estimates of how many users they can expect to reach given their targeting criteria and budget, without actually running the ad. 
As shown in Panel~3 of Figure~\ref{fig:customs}, we first target each Custom Audience separately without other criteria and use the API to measure the number of active users in each.
Then, we narrow down our targeted audience to only these users in each audience who share a given interest. 
We use the same API to measure the number of such users.
We then divide this number by the total number of users in each audience, see Panel~4 of Figure~\ref{fig:customs}. 
For example, we can create an ad with a daily budget of \$1M and target the Custom Audience that contains only the Republican voters. 
The interface then uses the \texttt{delivery\_estimate} endpoint of the Graph API to obtain an estimate, in this case 903,884 unique accounts.
We use this number as the estimate of the total number of weekly active users among our Republican list, $N_R$.
Using the same API endpoint and the conjunctive combinations feature we can also obtain the estimated number of Republicans in our custom audience interested in `Ted Nugent', $N_R^i$, in this case 111,374, or 12.3\% of the Republican audience.
We use the \texttt{delivery\_estimate} API to obtain these estimates for all 19K interests for all five custom audiences we created.

As a final step, we can compare the relative popularity of any interest $i$ between two non-overlapping audiences $A$ and $B$ (for example Republican vs Democratic, white vs Black, etc.) to estimate that interests' skew (See Panel 5 in Fig~\ref{fig:customs}).
We do so using the following equation:
\begin{equation}
    S^i_{AB} = \frac{\frac{N_A^i}{N_A}-\frac{N_B^i}{N_B}}{\frac{N^i_A}{N_A}+\frac{N^i_B}{N_B}},
    \label{eq:skew_def}
\end{equation}
where $N_A^i$ is the estimated number of weekly active users in group $A$ for whom Facebook inferred an interest in $i$ and $N_A$ is the total estimate of weekly active users in group $A$.
Similarly to measures we described in Section~\ref{sec:related-skews}, it is continuous from $-1$ to $1$. 

For example, the API estimates that up to 111,374, or 12.3\% of Republicans in our custom audience are interested in a country musician Ted Nugent. 
At the same time 43,085 out of 822,108, or 5.2\% Democrats in our custom audience are interested in `Ted Nugent'.
Therefore, the skew of `Ted Nugent' on the Republican vs. Democrat spectrum is $(12.3\%-5.2\%)/(12.3\%+5.2\%)=0.40$, i.e., it is skewing Republican and we can say that interest in `Ted Nugent' is a proxy for being Republican.

\begin{figure*}[t!]
    \includegraphics[width=1\linewidth]{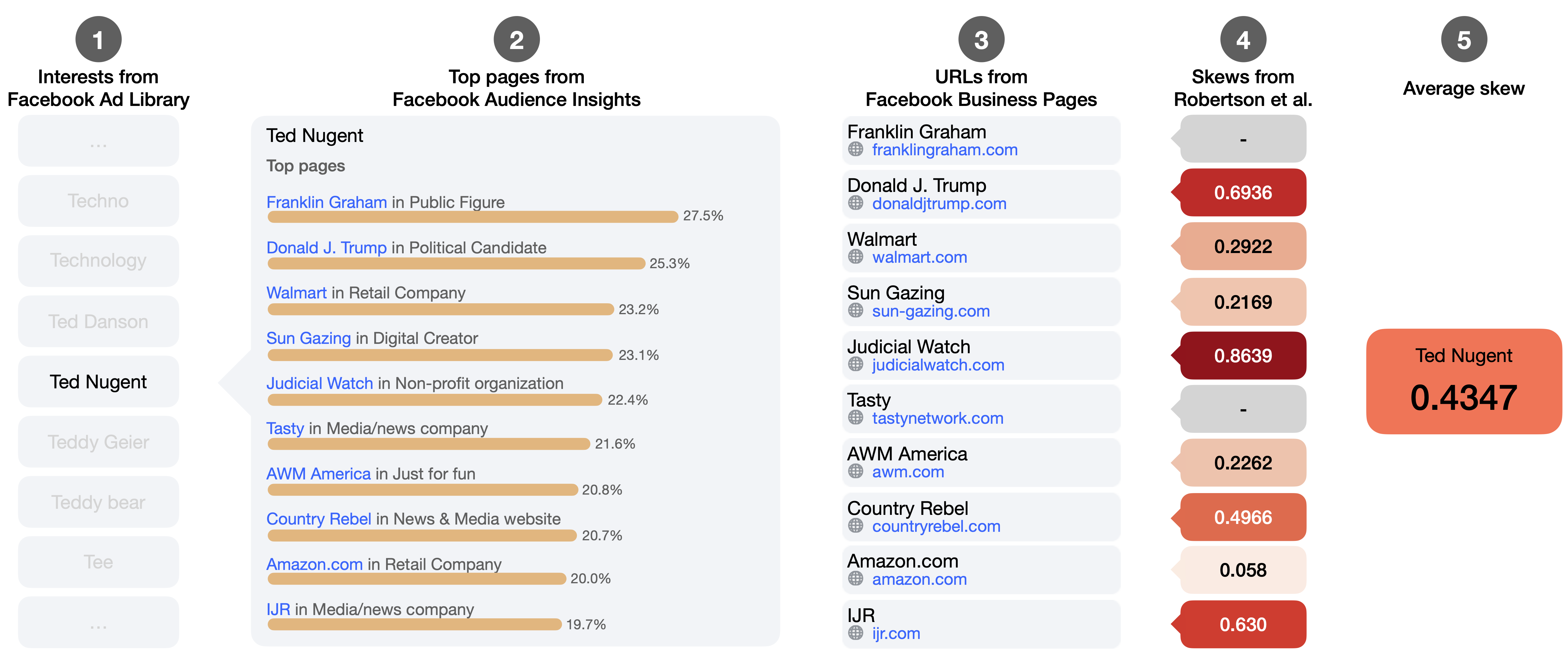}
    \caption{Estimating the skew of targeting interests using the Audience Insights data. \\ Step 1: Obtain the list of interests used by advertisers from the Facebook Ad Library. \\ Step 2: Obtain Facebook Pages popular among the users associated with a given interest. \\ Step 3: Obtain the domain of the external URL self-reported by that Facebook Page. \\ Step 4: Look-up the audience skew of each domain. \\ Step 5: Average the audience skews of known domains and assign it to the interest.}
    \label{fig:insights}
\end{figure*}

We note that this approach allows for continuous monitoring of the skew of different interests -- Custom Audiences can be continuously re-created with the newest releases of the voter records, targeting criteria that are actually used by advertisers can be obtained from the Ad Library, and the API can be then queried again to measure the prevalence of each interest in each Custom Audience.

\subsection{Estimating political skew through skew of external websites}
Our \textbf{second approach} is summarized in Figure~\ref{fig:insights}. 
It involves estimating the audience skew of a particular interest by leveraging known audience skews of websites that are popular among users who share that interest.
After obtaining the list of interests as described in the previous section, we visited the Facebook Audience Insights page for each interest. 
This tool allows any logged in user to enter a targeting specification (in our case: one interest) and to learn the characteristics of users who share that interest: age, gender, and geographical distributions, as well as their most popularly liked Facebook Pages, as shown in Panel~2 of Figure~\ref{fig:insights}.
Note that these Facebook Pages are {\em not} of individual users; instead they correspond to businesses, prominent figures, and celebrities.
Then, we visited each of these pages and collected the external URL that corresponds to that entity's website outside of Facebook, see Panel~3 of Figure~\ref{fig:insights}.
We find that the Audience Insights for 99\% of interests mention between seven and ten Facebook pages with associated domains.
To minimize the strain on Facebook's server and increase the speed of data collection, we collect the Pages data through the Graph API, rather than by visiting the Insights Page for every interest. 
We repeat the process for all unique interests in the Political Ad Targeting Dataset and thus obtain a mapping between each interest and the external URLs popular among users who share that interest.

To estimate the audience skew of an interest given the URLs associated with it, we need a measure of each URL's audience skew (see Panel~4 of Figure~\ref{fig:insights}).
We use the dataset published by \citet{robertson2018bias}.\footnote{The dataset can be downloaded from \url{https://github.com/gitronald/domains/}} 
Each of the domains in the dataset has a score between -1 (only Democratic voters linked to that domain in their tweets) through 0 (Democratic and Republican voters linked to that domain equally often) to 1 (only Republican voters linked to that domain).
We note that the dataset is far from complete: the scores have not been updated since February 2019 and the skew of particular domains might have changed over time, some domains may no longer exist, and new domains appeared.
Regardless, we believe the available scores are a good first approximation of the current reality, and we do have the bias estimate for 68\% of domain mentions.\footnote{The Robertson dataset covers only 21\% of unique domains from the Page Insights dataset, but missing domains are more likely to be less popular, which translates to higher empirical coverage.} Finally, as shown in Panel~5 of Figure~\ref{fig:insights}, we take the average audience skew of all URLs associated with an interest as the estimate of that interests' audience skew.

\subsection{Establishing the partisanship on political advertisers. }
\label{sec:whotargetsme}
To understand the political alignment of the advertisers present in the Ad Library data, we turned to an ad transparency non-profit called WhoTargetsMe.\footnote{See: \url{https://whotargets.me/en/projects/}.}
This organization allowed us to use a dataset which identifies the partisanship of 7,926 advertisers who ran political ads targeting users in the U.S.
Based on their identity and advertised content each advertiser is manually assigned one of the following nine labels: \texttt{Non} (for non-political advertisers), \texttt{GOP}, \texttt{Dems} (affiliated with the Republican Party or the Democratic Party, respectively), \texttt{R-PACs}, \texttt{D-PACs} (Political Action Committees associated with each party), \texttt{Conservative}, \texttt{Progressive} (advertising a right- or left-leaning point of view without official party affiliation), \texttt{Independent}, or \texttt{Other}. 
For the purpose of the analysis we collapse the labels into ``Conservatives'', ``Progressives'', and ``Other'', as described in Table~\ref{tab:whotargetsme}.

\begin{table}
    \small
    \centering
    \begin{tabular}{rlr}
    \textbf{Grouping} & \textbf{Label} & \textbf{Count} \\ \hline
    \multirow{4}*{Conservatives}    & \texttt{GOP}          & 1,289 \\
                                    & \texttt{R-PACs}       & 372 \\
                                    & \texttt{Conservative} & 221 \\
                                    & \textbf{Total}        & \textbf{1,882} \vspace{.5em}\\ 
                                    
    \multirow{4}*{Progressives}     & \texttt{Dems}         & 1,235 \\
                                    & \texttt{D-PACs}       & 572 \\
                                    & \texttt{Progressive}  & 148 \\
                                    & \textbf{Total}        & \textbf{1,955} \vspace{.5em}\\
                                    
    \multirow{4}*{Other}            & \texttt{Non}          & 3,924 \\
                                    & \texttt{Other}        & 137 \\
                                    & \texttt{Independent}  & 27 \\
                                    & \textbf{Total}        & \textbf{4,088} \\ \hline
                                    & \textbf{Total}        & \textbf{7,926}
    \end{tabular}
    \caption{Label breakdown in the advertiser affiliation dataset from WhoTargetsMe.}
    \label{tab:whotargetsme}
    
\end{table}

\section{Results}
We now present results of investigations into each of the research questions we set out to answer.
In short, we find that numerous interests that still remain targetable serve as effective proxies for political affiliation and race, and are actively exploited by advertisers on both sides of the political spectrum.

\begin{figure*}[t!]
    \includegraphics[width=1\linewidth]{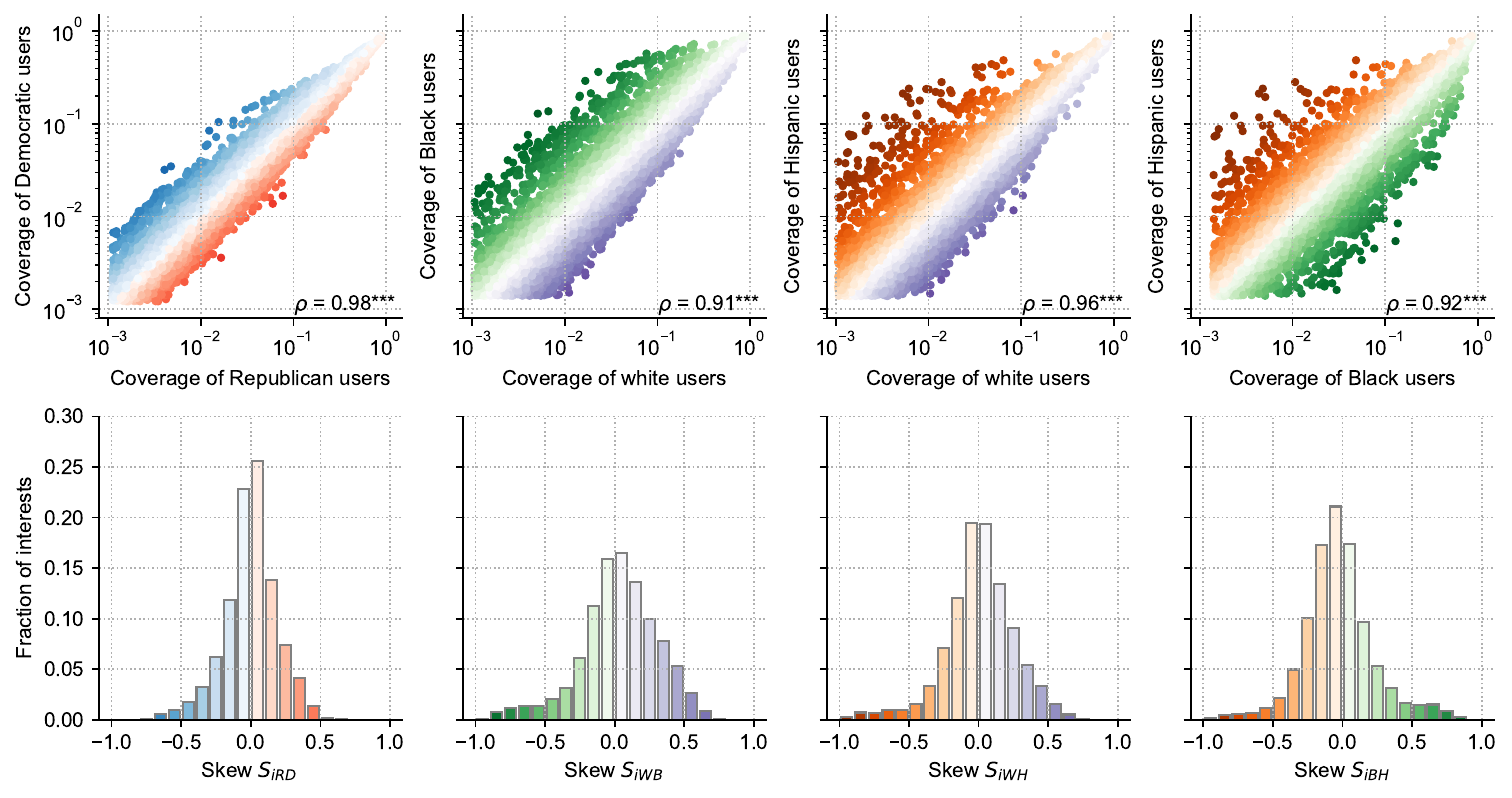}
    \caption{The coverage of interests between populations are highly correlated. Still, a non-trivial fraction of interests reveal large skews (color-coded in the figures) and could be used to selectively target populations that are not explicitly targetable.}
    \label{fig:skew}
\end{figure*}

\subsection{The existing targeting options can be used as proxies for political leaning and race.}
Figure~\ref{fig:skew} illustrates the prevalence of skews in the Political Targeting  dataset.
In the top panel figures each dot represents an interest. 
The color of the dots in the top panel encodes the skew: the more purple interests skew towards the population described in the $y$-axis and the interests in green skew towards the population described on the $x$-axis.
The figures show that, in general, coverage of interests is highly correlated between disjoint groups of users, i.e., if interest $i$ is popular among users in group $A$, it is likely also popular among users in group $B$. 
Among the comparisons, we see that Democrats and Republicans are the most similar to each other, with the coverage correlation of $\rho=0.98$ and $p_{val}<0.001$. White and Black users are the most dissimilar but still show a very high correlation at $\rho=0.91$, $p_{val}<0.001$. 
The histograms in the bottom panel show the distribution of skews and confirm the results measured using correlation: a small fraction of interests skews strongly towards Democrats or Republicans, but there exist many interests that skew strongly towards Black and Hispanic users.

Importantly, we do not observe interests that would constitute \textit{perfect} proxies. There are no clear outliers in the top left or bottom right in any of the plots---these would represent interests that fully cover one group without including any users from the other. 
Nevertheless, as we show in the following sections, it does not mean that the advertisers do not use the available \textit{imperfect} proxies.

\subsubsection{Comparison between the page-based and voter-based metrics}
\begin{figure}[t!]
    \includegraphics[width=.6\linewidth]{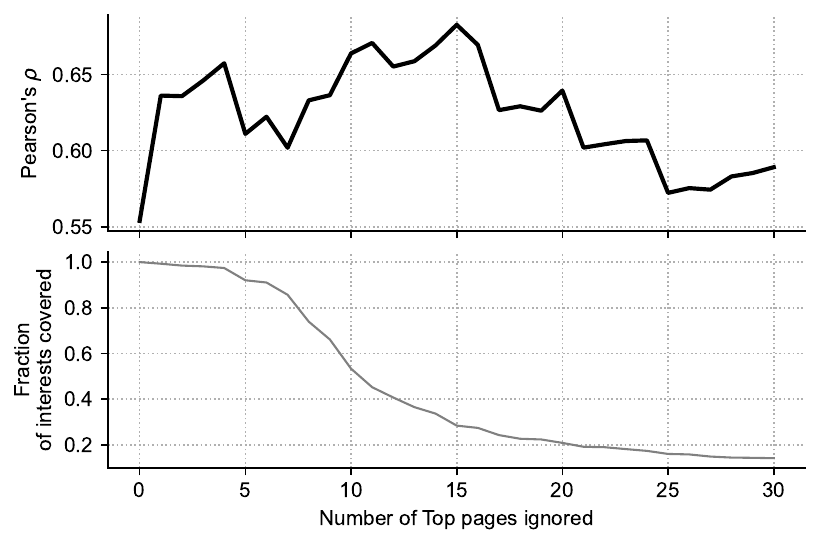}
    \caption{The skews calculated by audience and by page are correlated}
    \label{fig:corr_coverage}
\end{figure}

Our two metrics attempt to quantify the same phenomenon using disjoint methods and information.
We expect their scores to be different but still highly correlated, and they are indeed, at Pearson's $\rho=0.51$ and $p_{val}<0.001$.
We also recognize that the most common top domains may carry little information as they are associated with the majority of interests.
For example, Walmart appears among the Top Pages for 90\% of all interests.
We find that as we remove top domains, the coverage of interest for which we can estimate political skew drops, but the correlation between the page-based, and voter-based skew metric grows, see Figure~\ref{fig:corr_coverage}.
When the top four domains are dropped, the correlation between the page-based and the voter-based skew metrics achieves $\rho=0.66$ and $p_{val}<0.001$, while retaining estimates for 97\% of interests.
Beyond this point the coverage drops significantly without much gain in correlation.

We introduced the page-based skew measure as a proof of concept, without fine-tuning,
to demonstrate that even when voter records are not available (which is the case in many countries outside of the U.S.), the skew of interests could be approximated using a third-party domain-based skew lookup.
Further fine-tuning of this page-based skew measure is, of course, possible, including through machine-learning based analyses. We leave the exploration of these approaches to future work. 

For the remainder of the paper we focus on the voter-based skew measure.

\begin{table*}
    \small
    \centering
    \scalebox{0.78}{
    \begin{tabular}{rrlccccc}
    Interest & Spend (exclusion) & Spend (inclusion) &  Political & $S_{RD}$ & $S_{WB}$ & $S_{WH}$ & $S_{BH}$\\ \hline
Politics & 0.0 M \tikz[baseline=0ex]\filldraw[lightgray] (0, 0) rectangle (0.000cm,1.5ex); & \tikz[baseline=0ex]\filldraw[lightgray] (0, 0) rectangle (1.492cm,1.5ex); 6.6 M & \cmark & \cellcolor[rgb]{1.00, 0.92, 0.89}0.06 & \cellcolor[rgb]{0.94, 0.93, 0.96}0.13 & \cellcolor[rgb]{0.94, 0.93, 0.96}0.12 & \cellcolor[rgb]{1.00, 0.96, 0.92}-0.01 \\
Community issues & 0.0 M \tikz[baseline=0ex]\filldraw[lightgray] (0, 0) rectangle (0.006cm,1.5ex); & \tikz[baseline=0ex]\filldraw[lightgray] (0, 0) rectangle (1.401cm,1.5ex); 6.2 M & \cmark & \cellcolor[rgb]{1.00, 0.94, 0.91}0.04 & \cellcolor[rgb]{0.95, 0.98, 0.94}-0.03 & \cellcolor[rgb]{0.96, 0.95, 0.97}0.08 & \cellcolor[rgb]{0.91, 0.96, 0.89}0.12 \\
Voting & 0.0 M \tikz[baseline=0ex]\filldraw[lightgray] (0, 0) rectangle (0.000cm,1.5ex); & \tikz[baseline=0ex]\filldraw[lightgray] (0, 0) rectangle (1.091cm,1.5ex); 4.8 M & \cmark & \cellcolor[rgb]{0.97, 0.98, 1.00}-0.00 & \cellcolor[rgb]{0.94, 0.93, 0.96}0.12 & \cellcolor[rgb]{0.86, 0.86, 0.92}0.24 & \cellcolor[rgb]{0.90, 0.96, 0.88}0.12 \\
Social change & 0.2 M \tikz[baseline=0ex]\filldraw[lightgray] (0, 0) rectangle (0.046cm,1.5ex); & \tikz[baseline=0ex]\filldraw[lightgray] (0, 0) rectangle (0.932cm,1.5ex); 4.1 M & \cmark & \cellcolor[rgb]{0.91, 0.95, 0.98}-0.07 & \cellcolor[rgb]{0.86, 0.95, 0.84}-0.17 & \cellcolor[rgb]{0.97, 0.97, 0.98}0.04 & \cellcolor[rgb]{0.82, 0.93, 0.80}0.21 \\
Politics and social issues & 0.0 M \tikz[baseline=0ex]\filldraw[lightgray] (0, 0) rectangle (0.000cm,1.5ex); & \tikz[baseline=0ex]\filldraw[lightgray] (0, 0) rectangle (0.916cm,1.5ex); 4.0 M & \cmark & \cellcolor[rgb]{1.00, 0.94, 0.91}0.03 & \cellcolor[rgb]{0.98, 0.97, 0.98}0.04 & \cellcolor[rgb]{0.97, 0.96, 0.98}0.06 & \cellcolor[rgb]{0.96, 0.98, 0.95}0.03 \\
Joe Rogan & 3.7 M \tikz[baseline=0ex]\filldraw[lightgray] (0, 0) rectangle (0.846cm,1.5ex); & \tikz[baseline=0ex]\filldraw[lightgray] (0, 0) rectangle (0.064cm,1.5ex); 0.3 M & & \cellcolor[rgb]{1.00, 0.89, 0.84}0.10 & \cellcolor[rgb]{0.98, 0.97, 0.98}0.03 & \cellcolor[rgb]{0.97, 0.97, 0.98}0.04 & \cellcolor[rgb]{0.97, 0.99, 0.96}0.01 \\
Activism & 0.0 M \tikz[baseline=0ex]\filldraw[lightgray] (0, 0) rectangle (0.000cm,1.5ex); & \tikz[baseline=0ex]\filldraw[lightgray] (0, 0) rectangle (0.852cm,1.5ex); 3.7 M & \cmark & \cellcolor[rgb]{0.96, 0.98, 1.00}-0.02 & \cellcolor[rgb]{0.92, 0.97, 0.90}-0.10 & \cellcolor[rgb]{0.96, 0.96, 0.98}0.07 & \cellcolor[rgb]{0.86, 0.95, 0.84}0.17 \\
Kid Rock & 3.7 M \tikz[baseline=0ex]\filldraw[lightgray] (0, 0) rectangle (0.838cm,1.5ex); & \tikz[baseline=0ex]\filldraw[lightgray] (0, 0) rectangle (0.009cm,1.5ex); 0.0 M & & \cellcolor[rgb]{0.99, 0.75, 0.65}0.24 & \cellcolor[rgb]{0.68, 0.68, 0.83}0.43 & \cellcolor[rgb]{0.65, 0.63, 0.80}0.47 & \cellcolor[rgb]{0.95, 0.98, 0.93}0.05 \\
Charity and causes & 0.0 M \tikz[baseline=0ex]\filldraw[lightgray] (0, 0) rectangle (0.001cm,1.5ex); & \tikz[baseline=0ex]\filldraw[lightgray] (0, 0) rectangle (0.841cm,1.5ex); 3.7 M & \cmark & \cellcolor[rgb]{1.00, 0.95, 0.92}0.02 & \cellcolor[rgb]{0.99, 0.98, 0.99}0.01 & \cellcolor[rgb]{0.96, 0.95, 0.97}0.08 & \cellcolor[rgb]{0.93, 0.97, 0.92}0.07 \\
Social movement & 0.1 M \tikz[baseline=0ex]\filldraw[lightgray] (0, 0) rectangle (0.016cm,1.5ex); & \tikz[baseline=0ex]\filldraw[lightgray] (0, 0) rectangle (0.708cm,1.5ex); 3.1 M & \cmark & \cellcolor[rgb]{0.91, 0.95, 0.98}-0.07 & \cellcolor[rgb]{0.90, 0.96, 0.88}-0.12 & \cellcolor[rgb]{0.97, 0.96, 0.98}0.06 & \cellcolor[rgb]{0.85, 0.94, 0.82}0.18 \\
Ted Nugent & 3.0 M \tikz[baseline=0ex]\filldraw[lightgray] (0, 0) rectangle (0.686cm,1.5ex); & \tikz[baseline=0ex]\filldraw[lightgray] (0, 0) rectangle (0.018cm,1.5ex); 0.1 M & & \cellcolor[rgb]{0.99, 0.54, 0.41}0.40 & \cellcolor[rgb]{0.59, 0.57, 0.77}0.53 & \cellcolor[rgb]{0.64, 0.63, 0.80}0.48 & \cellcolor[rgb]{1.00, 0.93, 0.86}-0.07 \\
Current events & 0.0 M \tikz[baseline=0ex]\filldraw[lightgray] (0, 0) rectangle (0.000cm,1.5ex); & \tikz[baseline=0ex]\filldraw[lightgray] (0, 0) rectangle (0.690cm,1.5ex); 3.0 M & & \cellcolor[rgb]{1.00, 0.94, 0.91}0.03 & \cellcolor[rgb]{0.99, 0.98, 0.99}0.01 & \cellcolor[rgb]{0.96, 0.95, 0.97}0.08 & \cellcolor[rgb]{0.93, 0.97, 0.91}0.07 \\
NPR & 0.7 M \tikz[baseline=0ex]\filldraw[lightgray] (0, 0) rectangle (0.150cm,1.5ex); & \tikz[baseline=0ex]\filldraw[lightgray] (0, 0) rectangle (0.518cm,1.5ex); 2.3 M & & \cellcolor[rgb]{0.87, 0.92, 0.97}-0.13 & \cellcolor[rgb]{0.90, 0.90, 0.94}0.18 & \cellcolor[rgb]{0.90, 0.90, 0.94}0.19 & \cellcolor[rgb]{0.97, 0.99, 0.96}0.01 \\
Volunteering & 0.0 M \tikz[baseline=0ex]\filldraw[lightgray] (0, 0) rectangle (0.000cm,1.5ex); & \tikz[baseline=0ex]\filldraw[lightgray] (0, 0) rectangle (0.640cm,1.5ex); 2.8 M & & \cellcolor[rgb]{1.00, 0.90, 0.85}0.09 & \cellcolor[rgb]{0.94, 0.93, 0.96}0.12 & \cellcolor[rgb]{0.91, 0.91, 0.95}0.17 & \cellcolor[rgb]{0.94, 0.98, 0.93}0.05 \\
Hunting & 1.3 M \tikz[baseline=0ex]\filldraw[lightgray] (0, 0) rectangle (0.300cm,1.5ex); & \tikz[baseline=0ex]\filldraw[lightgray] (0, 0) rectangle (0.334cm,1.5ex); 1.5 M & & \cellcolor[rgb]{0.99, 0.82, 0.75}0.18 & \cellcolor[rgb]{0.93, 0.92, 0.96}0.15 & \cellcolor[rgb]{0.88, 0.88, 0.93}0.22 & \cellcolor[rgb]{0.93, 0.97, 0.91}0.07 \\
Local news & 0.0 M \tikz[baseline=0ex]\filldraw[lightgray] (0, 0) rectangle (0.000cm,1.5ex); & \tikz[baseline=0ex]\filldraw[lightgray] (0, 0) rectangle (0.608cm,1.5ex); 2.7 M & & \cellcolor[rgb]{1.00, 0.91, 0.87}0.08 & \cellcolor[rgb]{0.97, 0.96, 0.98}0.05 & \cellcolor[rgb]{0.90, 0.89, 0.94}0.19 & \cellcolor[rgb]{0.89, 0.96, 0.87}0.14 \\
Election & 0.0 M \tikz[baseline=0ex]\filldraw[lightgray] (0, 0) rectangle (0.000cm,1.5ex); & \tikz[baseline=0ex]\filldraw[lightgray] (0, 0) rectangle (0.576cm,1.5ex); 2.5 M & \cmark & \cellcolor[rgb]{1.00, 0.89, 0.84}0.12 & \cellcolor[rgb]{0.80, 0.80, 0.89}0.31 & \cellcolor[rgb]{0.85, 0.85, 0.92}0.26 & \cellcolor[rgb]{1.00, 0.94, 0.88}-0.05 \\
Local government & 0.0 M \tikz[baseline=0ex]\filldraw[lightgray] (0, 0) rectangle (0.000cm,1.5ex); & \tikz[baseline=0ex]\filldraw[lightgray] (0, 0) rectangle (0.534cm,1.5ex); 2.3 M & \cmark & \cellcolor[rgb]{1.00, 0.89, 0.84}0.11 & \cellcolor[rgb]{0.86, 0.86, 0.93}0.24 & \cellcolor[rgb]{0.96, 0.95, 0.97}0.08 & \cellcolor[rgb]{1.00, 0.88, 0.76}-0.16 \\
Duck Dynasty & 1.6 M \tikz[baseline=0ex]\filldraw[lightgray] (0, 0) rectangle (0.371cm,1.5ex); & \tikz[baseline=0ex]\filldraw[lightgray] (0, 0) rectangle (0.162cm,1.5ex); 0.7 M & & \cellcolor[rgb]{0.99, 0.49, 0.36}0.44 & \cellcolor[rgb]{0.66, 0.65, 0.81}0.46 & \cellcolor[rgb]{0.70, 0.69, 0.84}0.42 & \cellcolor[rgb]{1.00, 0.94, 0.88}-0.05 \\
NASCAR & 1.7 M \tikz[baseline=0ex]\filldraw[lightgray] (0, 0) rectangle (0.378cm,1.5ex); & \tikz[baseline=0ex]\filldraw[lightgray] (0, 0) rectangle (0.125cm,1.5ex); 0.6 M & & \cellcolor[rgb]{0.99, 0.72, 0.62}0.26 & \cellcolor[rgb]{0.86, 0.86, 0.93}0.24 & \cellcolor[rgb]{0.85, 0.85, 0.92}0.26 & \cellcolor[rgb]{0.96, 0.98, 0.95}0.02 \\
Parents  & 0.0 M \tikz[baseline=0ex]\filldraw[lightgray] (0, 0) rectangle (0.000cm,1.5ex); & \tikz[baseline=0ex]\filldraw[lightgray] (0, 0) rectangle (0.491cm,1.5ex); 2.2 M & & \cellcolor[rgb]{1.00, 0.88, 0.82}0.13 & \cellcolor[rgb]{0.98, 0.97, 0.98}0.03 & \cellcolor[rgb]{0.90, 0.90, 0.94}0.18 & \cellcolor[rgb]{0.87, 0.95, 0.85}0.15 \\
Fox News & 1.6 M \tikz[baseline=0ex]\filldraw[lightgray] (0, 0) rectangle (0.358cm,1.5ex); & \tikz[baseline=0ex]\filldraw[lightgray] (0, 0) rectangle (0.099cm,1.5ex); 0.4 M & & \cellcolor[rgb]{.9, .9, .9}- & \cellcolor[rgb]{.9, .9, .9}- & \cellcolor[rgb]{.9, .9, .9}- & \cellcolor[rgb]{.9, .9, .9}- \\
Newspapers & 0.0 M \tikz[baseline=0ex]\filldraw[lightgray] (0, 0) rectangle (0.006cm,1.5ex); & \tikz[baseline=0ex]\filldraw[lightgray] (0, 0) rectangle (0.445cm,1.5ex); 2.0 M & & \cellcolor[rgb]{1.00, 0.92, 0.89}0.06 & \cellcolor[rgb]{0.93, 0.92, 0.96}0.14 & \cellcolor[rgb]{0.98, 0.97, 0.98}0.03 & \cellcolor[rgb]{1.00, 0.91, 0.83}-0.10 \\
Nonprofit organization & 0.0 M \tikz[baseline=0ex]\filldraw[lightgray] (0, 0) rectangle (0.000cm,1.5ex); & \tikz[baseline=0ex]\filldraw[lightgray] (0, 0) rectangle (0.440cm,1.5ex); 1.9 M & & \cellcolor[rgb]{1.00, 0.91, 0.87}0.08 & \cellcolor[rgb]{0.95, 0.94, 0.97}0.11 & \cellcolor[rgb]{0.91, 0.91, 0.95}0.17 & \cellcolor[rgb]{0.94, 0.98, 0.92}0.06 \\
Breaking news & 0.1 M \tikz[baseline=0ex]\filldraw[lightgray] (0, 0) rectangle (0.019cm,1.5ex); & \tikz[baseline=0ex]\filldraw[lightgray] (0, 0) rectangle (0.410cm,1.5ex); 1.8 M & & \cellcolor[rgb]{1.00, 0.93, 0.90}0.05 & \cellcolor[rgb]{0.97, 0.96, 0.98}0.05 & \cellcolor[rgb]{0.94, 0.94, 0.96}0.11 & \cellcolor[rgb]{0.93, 0.97, 0.92}0.06 \\
Fishing & 0.3 M \tikz[baseline=0ex]\filldraw[lightgray] (0, 0) rectangle (0.060cm,1.5ex); & \tikz[baseline=0ex]\filldraw[lightgray] (0, 0) rectangle (0.349cm,1.5ex); 1.5 M & & \cellcolor[rgb]{0.99, 0.83, 0.76}0.17 & \cellcolor[rgb]{0.91, 0.91, 0.95}0.16 & \cellcolor[rgb]{0.93, 0.92, 0.96}0.14 & \cellcolor[rgb]{1.00, 0.95, 0.90}-0.02 \\
Government & 0.0 M \tikz[baseline=0ex]\filldraw[lightgray] (0, 0) rectangle (0.000cm,1.5ex); & \tikz[baseline=0ex]\filldraw[lightgray] (0, 0) rectangle (0.395cm,1.5ex); 1.7 M & \cmark & \cellcolor[rgb]{0.99, 0.85, 0.79}0.15 & \cellcolor[rgb]{0.92, 0.91, 0.95}0.15 & \cellcolor[rgb]{0.92, 0.91, 0.95}0.16 & \cellcolor[rgb]{0.96, 0.99, 0.96}0.01 \\
Sustainable energy & 0.1 M \tikz[baseline=0ex]\filldraw[lightgray] (0, 0) rectangle (0.017cm,1.5ex); & \tikz[baseline=0ex]\filldraw[lightgray] (0, 0) rectangle (0.368cm,1.5ex); 1.6 M & & \cellcolor[rgb]{1.00, 0.93, 0.90}0.04 & \cellcolor[rgb]{0.92, 0.91, 0.95}0.15 & \cellcolor[rgb]{0.98, 0.98, 0.99}0.02 & \cellcolor[rgb]{1.00, 0.90, 0.80}-0.13 \\
Election day & 0.0 M \tikz[baseline=0ex]\filldraw[lightgray] (0, 0) rectangle (0.000cm,1.5ex); & \tikz[baseline=0ex]\filldraw[lightgray] (0, 0) rectangle (0.373cm,1.5ex); 1.6 M & \cmark & \cellcolor[rgb]{0.94, 0.96, 0.99}-0.04 & \cellcolor[rgb]{0.93, 0.97, 0.92}-0.07 & \cellcolor[rgb]{1.00, 0.91, 0.83}-0.11 & \cellcolor[rgb]{1.00, 0.94, 0.89}-0.04 \\
Roseanne Barr & 1.5 M \tikz[baseline=0ex]\filldraw[lightgray] (0, 0) rectangle (0.341cm,1.5ex); & \tikz[baseline=0ex]\filldraw[lightgray] (0, 0) rectangle (0.007cm,1.5ex); 0.0 M & & \cellcolor[rgb]{0.94, 0.97, 0.99}-0.03 & \cellcolor[rgb]{0.87, 0.87, 0.93}0.22 & \cellcolor[rgb]{0.89, 0.89, 0.94}0.20 & \cellcolor[rgb]{1.00, 0.95, 0.90}-0.03 \\
Community and Social Services & 0.0 M \tikz[baseline=0ex]\filldraw[lightgray] (0, 0) rectangle (0.000cm,1.5ex); & \tikz[baseline=0ex]\filldraw[lightgray] (0, 0) rectangle (0.346cm,1.5ex); 1.5 M & \cmark & \cellcolor[rgb]{0.85, 0.91, 0.96}-0.16 & \cellcolor[rgb]{0.91, 0.90, 0.95}0.18 & \cellcolor[rgb]{0.92, 0.91, 0.95}0.16 & \cellcolor[rgb]{1.00, 0.95, 0.91}-0.02 \\
Love \& Hip Hop & 0.0 M \tikz[baseline=0ex]\filldraw[lightgray] (0, 0) rectangle (0.000cm,1.5ex); & \tikz[baseline=0ex]\filldraw[lightgray] (0, 0) rectangle (0.338cm,1.5ex); 1.5 M & & \cellcolor[rgb]{0.70, 0.83, 0.91}-0.31 & \cellcolor[rgb]{0.26, 0.67, 0.37}-0.62 & \cellcolor[rgb]{1.00, 0.93, 0.86}-0.07 & \cellcolor[rgb]{0.34, 0.71, 0.40}0.58 \\
Parents with early school-age children  & 0.0 M \tikz[baseline=0ex]\filldraw[lightgray] (0, 0) rectangle (0.000cm,1.5ex); & \tikz[baseline=0ex]\filldraw[lightgray] (0, 0) rectangle (0.337cm,1.5ex); 1.5 M & & \cellcolor[rgb]{1.00, 0.93, 0.90}0.05 & \cellcolor[rgb]{0.89, 0.89, 0.94}0.20 & \cellcolor[rgb]{0.94, 0.93, 0.96}0.13 & \cellcolor[rgb]{1.00, 0.93, 0.86}-0.07 \\
African-American literature & 0.0 M \tikz[baseline=0ex]\filldraw[lightgray] (0, 0) rectangle (0.000cm,1.5ex); & \tikz[baseline=0ex]\filldraw[lightgray] (0, 0) rectangle (0.335cm,1.5ex); 1.5 M & & \cellcolor[rgb]{0.67, 0.81, 0.90}-0.34 & \cellcolor[rgb]{0.24, 0.66, 0.36}-0.63 & \cellcolor[rgb]{1.00, 0.96, 0.92}-0.00 & \cellcolor[rgb]{0.24, 0.66, 0.36}0.63 \\
The Real Housewives of Atlanta & 0.0 M \tikz[baseline=0ex]\filldraw[lightgray] (0, 0) rectangle (0.000cm,1.5ex); & \tikz[baseline=0ex]\filldraw[lightgray] (0, 0) rectangle (0.332cm,1.5ex); 1.5 M & & \cellcolor[rgb]{0.80, 0.87, 0.94}-0.22 & \cellcolor[rgb]{0.44, 0.76, 0.45}-0.51 & \cellcolor[rgb]{0.98, 0.97, 0.99}0.03 & \cellcolor[rgb]{0.41, 0.75, 0.44}0.53 \\
Parents with preteens  & 0.0 M \tikz[baseline=0ex]\filldraw[lightgray] (0, 0) rectangle (0.000cm,1.5ex); & \tikz[baseline=0ex]\filldraw[lightgray] (0, 0) rectangle (0.331cm,1.5ex); 1.5 M & & \cellcolor[rgb]{1.00, 0.90, 0.85}0.10 & \cellcolor[rgb]{0.91, 0.90, 0.95}0.17 & \cellcolor[rgb]{0.91, 0.91, 0.95}0.16 & \cellcolor[rgb]{1.00, 0.96, 0.91}-0.01 \\
Clint Eastwood & 1.4 M \tikz[baseline=0ex]\filldraw[lightgray] (0, 0) rectangle (0.318cm,1.5ex); & \tikz[baseline=0ex]\filldraw[lightgray] (0, 0) rectangle (0.006cm,1.5ex); 0.0 M & & \cellcolor[rgb]{0.99, 0.64, 0.52}0.33 & \cellcolor[rgb]{0.56, 0.54, 0.75}0.56 & \cellcolor[rgb]{0.70, 0.70, 0.84}0.41 & \cellcolor[rgb]{0.99, 0.85, 0.71}-0.20 \\
Electoral reform & 0.0 M \tikz[baseline=0ex]\filldraw[lightgray] (0, 0) rectangle (0.000cm,1.5ex); & \tikz[baseline=0ex]\filldraw[lightgray] (0, 0) rectangle (0.322cm,1.5ex); 1.4 M & \cmark & \cellcolor[rgb]{1.00, 0.87, 0.81}0.14 & \cellcolor[rgb]{0.94, 0.93, 0.96}0.13 & \cellcolor[rgb]{0.78, 0.78, 0.88}0.33 & \cellcolor[rgb]{0.82, 0.93, 0.79}0.21 \\
Parents with teenagers  & 0.0 M \tikz[baseline=0ex]\filldraw[lightgray] (0, 0) rectangle (0.000cm,1.5ex); & \tikz[baseline=0ex]\filldraw[lightgray] (0, 0) rectangle (0.321cm,1.5ex); 1.4 M & & \cellcolor[rgb]{1.00, 0.88, 0.83}0.12 & \cellcolor[rgb]{0.99, 0.98, 0.99}0.00 & \cellcolor[rgb]{0.91, 0.90, 0.95}0.18 & \cellcolor[rgb]{0.86, 0.95, 0.84}0.17 \\
College grad & 0.3 M \tikz[baseline=0ex]\filldraw[lightgray] (0, 0) rectangle (0.073cm,1.5ex); & \tikz[baseline=0ex]\filldraw[lightgray] (0, 0) rectangle (0.243cm,1.5ex); 1.1 M & & \cellcolor[rgb]{1.00, 0.95, 0.93}0.02 & \cellcolor[rgb]{0.99, 0.98, 0.99}0.01 & \cellcolor[rgb]{0.99, 0.98, 0.99}0.01 & \cellcolor[rgb]{0.97, 0.99, 0.96}0.00 \\
Charitable organization & 0.0 M \tikz[baseline=0ex]\filldraw[lightgray] (0, 0) rectangle (0.000cm,1.5ex); & \tikz[baseline=0ex]\filldraw[lightgray] (0, 0) rectangle (0.311cm,1.5ex); 1.4 M & & \cellcolor[rgb]{1.00, 0.94, 0.91}0.03 & \cellcolor[rgb]{0.97, 0.96, 0.98}0.06 & \cellcolor[rgb]{0.92, 0.91, 0.95}0.15 & \cellcolor[rgb]{0.92, 0.97, 0.90}0.09 \\
Renewable energy & 0.1 M \tikz[baseline=0ex]\filldraw[lightgray] (0, 0) rectangle (0.030cm,1.5ex); & \tikz[baseline=0ex]\filldraw[lightgray] (0, 0) rectangle (0.275cm,1.5ex); 1.2 M & & \cellcolor[rgb]{1.00, 0.93, 0.90}0.04 & \cellcolor[rgb]{0.93, 0.92, 0.96}0.14 & \cellcolor[rgb]{1.00, 0.95, 0.91}-0.02 & \cellcolor[rgb]{1.00, 0.88, 0.77}-0.15 \\
Black-ish & 0.0 M \tikz[baseline=0ex]\filldraw[lightgray] (0, 0) rectangle (0.000cm,1.5ex); & \tikz[baseline=0ex]\filldraw[lightgray] (0, 0) rectangle (0.303cm,1.5ex); 1.3 M & & \cellcolor[rgb]{0.56, 0.76, 0.87}-0.41 & \cellcolor[rgb]{0.19, 0.60, 0.31}-0.69 & \cellcolor[rgb]{1.00, 0.91, 0.83}-0.10 & \cellcolor[rgb]{0.25, 0.67, 0.36}0.63 \\
News media & 0.1 M \tikz[baseline=0ex]\filldraw[lightgray] (0, 0) rectangle (0.019cm,1.5ex); & \tikz[baseline=0ex]\filldraw[lightgray] (0, 0) rectangle (0.275cm,1.5ex); 1.2 M & & \cellcolor[rgb]{1.00, 0.92, 0.89}0.06 & \cellcolor[rgb]{0.94, 0.98, 0.92}-0.06 & \cellcolor[rgb]{0.91, 0.90, 0.95}0.18 & \cellcolor[rgb]{0.80, 0.92, 0.77}0.23 \\
Philanthropy & 0.0 M \tikz[baseline=0ex]\filldraw[lightgray] (0, 0) rectangle (0.000cm,1.5ex); & \tikz[baseline=0ex]\filldraw[lightgray] (0, 0) rectangle (0.293cm,1.5ex); 1.3 M & & \cellcolor[rgb]{0.96, 0.98, 1.00}-0.01 & \cellcolor[rgb]{0.80, 0.92, 0.77}-0.23 & \cellcolor[rgb]{0.97, 0.96, 0.98}0.05 & \cellcolor[rgb]{0.76, 0.90, 0.73}0.27 \\
Elon Musk & 1.2 M \tikz[baseline=0ex]\filldraw[lightgray] (0, 0) rectangle (0.276cm,1.5ex); & \tikz[baseline=0ex]\filldraw[lightgray] (0, 0) rectangle (0.007cm,1.5ex); 0.0 M & & \cellcolor[rgb]{1.00, 0.87, 0.81}0.14 & \cellcolor[rgb]{0.95, 0.95, 0.97}0.09 & \cellcolor[rgb]{0.95, 0.95, 0.97}0.09 & \cellcolor[rgb]{1.00, 0.96, 0.92}-0.00 \\
Small business & 0.0 M \tikz[baseline=0ex]\filldraw[lightgray] (0, 0) rectangle (0.000cm,1.5ex); & \tikz[baseline=0ex]\filldraw[lightgray] (0, 0) rectangle (0.280cm,1.5ex); 1.2 M & & \cellcolor[rgb]{1.00, 0.93, 0.90}0.05 & \cellcolor[rgb]{0.93, 0.97, 0.92}-0.06 & \cellcolor[rgb]{0.98, 0.97, 0.98}0.04 & \cellcolor[rgb]{0.92, 0.97, 0.90}0.10 \\
Basketball Wives & 0.0 M \tikz[baseline=0ex]\filldraw[lightgray] (0, 0) rectangle (0.000cm,1.5ex); & \tikz[baseline=0ex]\filldraw[lightgray] (0, 0) rectangle (0.279cm,1.5ex); 1.2 M & & \cellcolor[rgb]{0.73, 0.84, 0.92}-0.28 & \cellcolor[rgb]{0.32, 0.70, 0.40}-0.58 & \cellcolor[rgb]{0.98, 0.98, 0.99}0.02 & \cellcolor[rgb]{0.30, 0.69, 0.39}0.60 \\
Hip hop music & 0.0 M \tikz[baseline=0ex]\filldraw[lightgray] (0, 0) rectangle (0.003cm,1.5ex); & \tikz[baseline=0ex]\filldraw[lightgray] (0, 0) rectangle (0.274cm,1.5ex); 1.2 M & & \cellcolor[rgb]{0.90, 0.94, 0.98}-0.09 & \cellcolor[rgb]{0.73, 0.89, 0.70}-0.30 & \cellcolor[rgb]{1.00, 0.94, 0.89}-0.05 & \cellcolor[rgb]{0.78, 0.91, 0.75}0.25 \\
Saturday Night Live & 0.5 M \tikz[baseline=0ex]\filldraw[lightgray] (0, 0) rectangle (0.118cm,1.5ex); & \tikz[baseline=0ex]\filldraw[lightgray] (0, 0) rectangle (0.157cm,1.5ex); 0.7 M & & \cellcolor[rgb]{0.93, 0.96, 0.99}-0.06 & \cellcolor[rgb]{0.98, 0.97, 0.99}0.03 & \cellcolor[rgb]{0.96, 0.96, 0.98}0.07 & \cellcolor[rgb]{0.95, 0.98, 0.94}0.04 \\
Barstool Sports & 0.6 M \tikz[baseline=0ex]\filldraw[lightgray] (0, 0) rectangle (0.140cm,1.5ex); & \tikz[baseline=0ex]\filldraw[lightgray] (0, 0) rectangle (0.134cm,1.5ex); 0.6 M & & \cellcolor[rgb]{0.99, 0.77, 0.68}0.22 & \cellcolor[rgb]{0.94, 0.94, 0.96}0.11 & \cellcolor[rgb]{0.87, 0.87, 0.93}0.23 & \cellcolor[rgb]{0.90, 0.96, 0.88}0.12 \\
Big-game hunting & 0.7 M \tikz[baseline=0ex]\filldraw[lightgray] (0, 0) rectangle (0.154cm,1.5ex); & \tikz[baseline=0ex]\filldraw[lightgray] (0, 0) rectangle (0.119cm,1.5ex); 0.5 M & & \cellcolor[rgb]{0.99, 0.78, 0.69}0.22 & \cellcolor[rgb]{0.86, 0.86, 0.92}0.25 & \cellcolor[rgb]{0.94, 0.94, 0.96}0.11 & \cellcolor[rgb]{1.00, 0.90, 0.80}-0.14 \\
The Shade Room & 0.0 M \tikz[baseline=0ex]\filldraw[lightgray] (0, 0) rectangle (0.000cm,1.5ex); & \tikz[baseline=0ex]\filldraw[lightgray] (0, 0) rectangle (0.273cm,1.5ex); 1.2 M & & \cellcolor[rgb]{0.14, 0.45, 0.72}-0.74 & \cellcolor[rgb]{0.00, 0.42, 0.17}-0.88 & \cellcolor[rgb]{0.99, 0.75, 0.52}-0.32 & \cellcolor[rgb]{0.10, 0.51, 0.24}0.78 \\
Soul music & 0.0 M \tikz[baseline=0ex]\filldraw[lightgray] (0, 0) rectangle (0.000cm,1.5ex); & \tikz[baseline=0ex]\filldraw[lightgray] (0, 0) rectangle (0.272cm,1.5ex); 1.2 M & & \cellcolor[rgb]{0.82, 0.89, 0.95}-0.20 & \cellcolor[rgb]{0.53, 0.80, 0.52}-0.45 & \cellcolor[rgb]{0.99, 0.98, 0.99}0.00 & \cellcolor[rgb]{0.53, 0.80, 0.52}0.45 \\
Deer hunting & 0.5 M \tikz[baseline=0ex]\filldraw[lightgray] (0, 0) rectangle (0.115cm,1.5ex); & \tikz[baseline=0ex]\filldraw[lightgray] (0, 0) rectangle (0.153cm,1.5ex); 0.7 M & & \cellcolor[rgb]{0.99, 0.85, 0.79}0.15 & \cellcolor[rgb]{0.88, 0.88, 0.93}0.21 & \cellcolor[rgb]{0.95, 0.94, 0.97}0.10 & \cellcolor[rgb]{1.00, 0.91, 0.82}-0.11 \\
Small business owners & 0.0 M \tikz[baseline=0ex]\filldraw[lightgray] (0, 0) rectangle (0.002cm,1.5ex); & \tikz[baseline=0ex]\filldraw[lightgray] (0, 0) rectangle (0.263cm,1.5ex); 1.2 M & & \cellcolor[rgb]{1.00, 0.91, 0.87}0.08 & \cellcolor[rgb]{0.95, 0.98, 0.94}-0.03 & \cellcolor[rgb]{0.99, 0.98, 0.99}0.01 & \cellcolor[rgb]{0.95, 0.98, 0.93}0.04 \\
Golf & 0.2 M \tikz[baseline=0ex]\filldraw[lightgray] (0, 0) rectangle (0.035cm,1.5ex); & \tikz[baseline=0ex]\filldraw[lightgray] (0, 0) rectangle (0.230cm,1.5ex); 1.0 M & & \cellcolor[rgb]{1.00, 0.88, 0.83}0.12 & \cellcolor[rgb]{0.92, 0.91, 0.95}0.15 & \cellcolor[rgb]{0.95, 0.94, 0.97}0.11 & \cellcolor[rgb]{1.00, 0.94, 0.89}-0.04 \\
Master's degree & 0.3 M \tikz[baseline=0ex]\filldraw[lightgray] (0, 0) rectangle (0.063cm,1.5ex); & \tikz[baseline=0ex]\filldraw[lightgray] (0, 0) rectangle (0.201cm,1.5ex); 0.9 M & & \cellcolor[rgb]{0.90, 0.94, 0.98}-0.09 & \cellcolor[rgb]{0.83, 0.93, 0.81}-0.20 & \cellcolor[rgb]{1.00, 0.91, 0.81}-0.12 & \cellcolor[rgb]{0.93, 0.97, 0.91}0.08 \\
Ayn Rand & 1.1 M \tikz[baseline=0ex]\filldraw[lightgray] (0, 0) rectangle (0.251cm,1.5ex); & \tikz[baseline=0ex]\filldraw[lightgray] (0, 0) rectangle (0.009cm,1.5ex); 0.0 M & & \cellcolor[rgb]{0.99, 0.70, 0.59}0.27 & \cellcolor[rgb]{0.76, 0.76, 0.87}0.35 & \cellcolor[rgb]{0.82, 0.82, 0.90}0.29 & \cellcolor[rgb]{1.00, 0.93, 0.86}-0.07 \\
Health care & 0.0 M \tikz[baseline=0ex]\filldraw[lightgray] (0, 0) rectangle (0.000cm,1.5ex); & \tikz[baseline=0ex]\filldraw[lightgray] (0, 0) rectangle (0.260cm,1.5ex); 1.1 M & \cmark & \cellcolor[rgb]{1.00, 0.95, 0.93}0.02 & \cellcolor[rgb]{0.96, 0.98, 0.95}-0.02 & \cellcolor[rgb]{0.98, 0.98, 0.99}0.02 & \cellcolor[rgb]{0.95, 0.98, 0.94}0.04 \\
    \end{tabular}}
    \caption{Top 60 Interest ordered by spend in USD. Political advertisers in the U.S.~spent millions of dollars targeting highly skewed interest circumventing the platform regulation.}
    \label{tab:big_table}
\end{table*}

\subsection{Real political advertisers spend millions of dollars on highly skewed interests. }
Table~\ref{tab:big_table} shows the interests used for targeting of political ads sorted by total amount spent, while distinguishing between targeting and exclusion.
The majority of these interests are related to the content and stated purpose of advertising, with `Politics', `Community Issues', `Voting', `Social change', and `Politics and social issues', `Activism', `Charity and causes', and `Social movement' in the top ten spots by spend.
We also observe that these interests are not skewed towards either political option or any ethnic group.

However, we also find that advertisers spent millions targeting and excluding interests that are not explicitly related to politics.
Users interested in `Joe Rogan', `Kid Rock', `Ted Nugent', `Duck Dynasty', `NASCAR', `Roseanne Barr', `Clint Eastwood', `Elon Musk', or `Ayn Rand' are often excluded from seeing political messaging.
The skew measures presented in the table show why advertisers would make such choices: all but two of these interests skew strongly Republican and white.
Notably, there is no skew information for the often excluded ``Fox News''. 
At the time of writing this paper, this targeting criteria referred to the people who were employed there, not merely interested in the topic.
Thus, that population was too small to obtain skew measures.

On the other hand, many interests that lean strongly towards Black voters---and at times refer to them directly---are used for including audiences for political ads, for example `Love \& Hip-Hop', `African-American literature', `Black-ish', `Basketball Wives', or `The Shade Room'.
Note that, with the exception of `NPR' (National Public Radio), the interests with stronger Democratic skews are  predominantly of interest to Black voters, who are largely Democratic, rather than Democrats at large. 
Finally, some interests ware used both for inclusion and exclusion: `Saturday Night Live', `Barstool Sports', `Big-game hunting', or `Deer hunting'.

The prevalence of use of interests with no apparent connection to politics, but often with high skews towards political or ethnic groups, indicates that advertisers are skillfully circumventing the platform's limitations on targeting.

\begin{figure}[t!]
    \includegraphics[width=.7\linewidth]{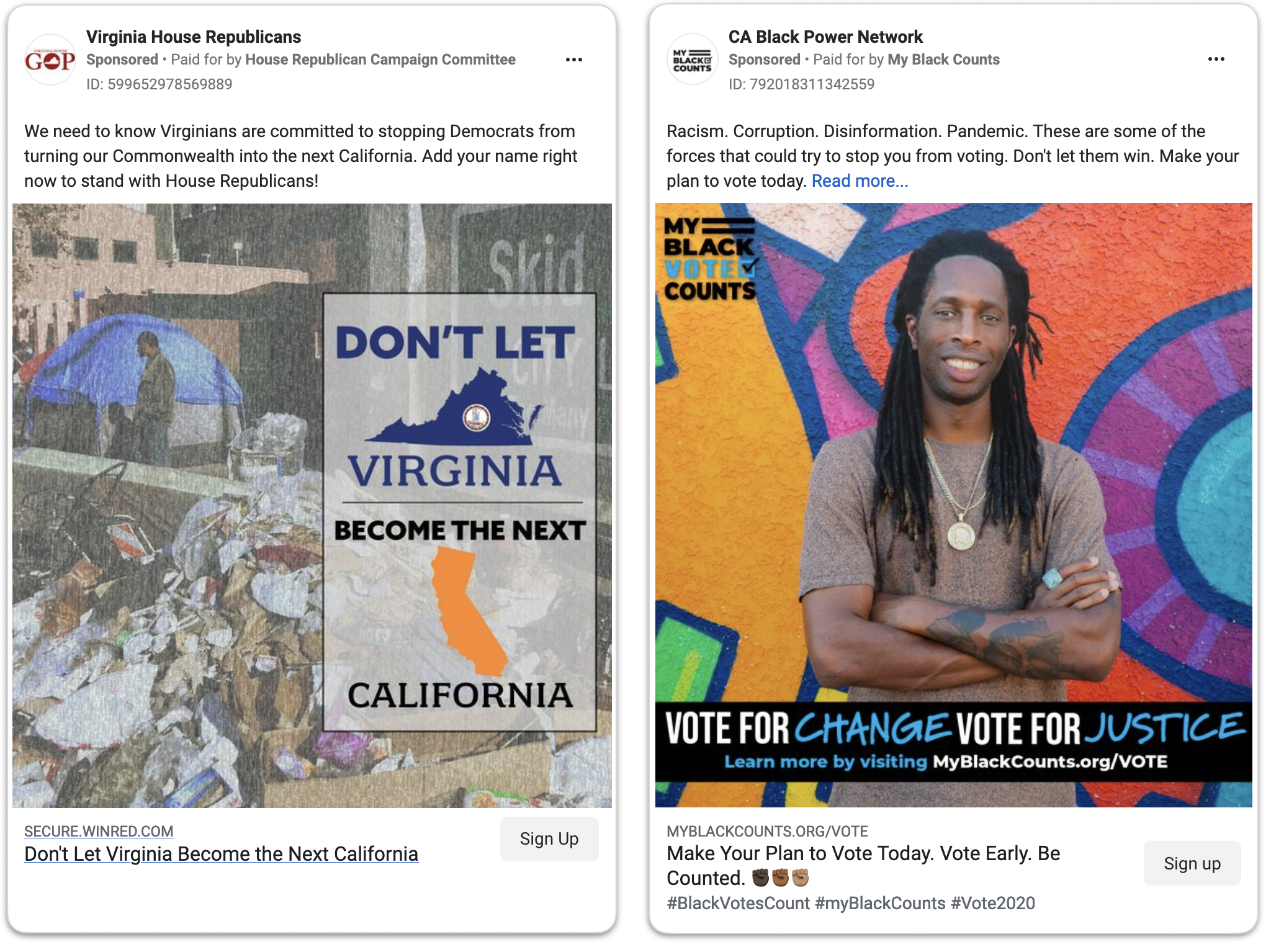}
    \caption{Examples of ads whose advertisers appear to use proxies. Virginia House Republicans (on the left) target users interested in (among others) NASCAR, Big-game hunting, and Deer hunting. On the right, CA Black Power Network targets users interested in The Shade Room, Basketball Wives, Love \& Hip Hop.}
    \label{fig:ad_examples}
\end{figure}

A reader familiar with the U.S.~media landscape likely understands the cultural background for most of the skewed interests in Table~\ref{tab:big_table}, especially that the majority of the interests skew in the direction they would anticipate.
Still, for the benefit of other readers we provide a short description.
Musicians Kid Rock and Ted Nugent, actors Clint Eastwood and Roseanne Barr, and billionaire Elon Musk have all been outspoken supporters of the Republican Party~\cite{rollingstone2016kidrock,cnn2018ted}.
Joe Rogan is a podcast host, who---while declining to align himself with either side of the political spectrum---tends to center right-wing voices~\cite{cnn2022rogan}.
Duck Dynasty is a reality show that features a family patriarch whose right to hold and promote anti-gay views was defended by prominent Republican politicians~\cite{usatoday2014duck}; 
NASCAR is a motor sport that, while it has no political presence, boasts a fan base that is predominantly older, white, male, and Republican~\cite{huff2013nascar}.
Ayn Rand is an author whose writing continues to inspire conservatives \cite{guardian_rand}.
Love \& Hip-Hop and Basketball Wives are reality TV series, showcasing hiphop and R'n'B musicians (the former), and women who are romantically involved with NBA players (the latter).
Black-ish is a sitcom with a largely African-American cast, and The Shade Room is a media company focused on celebrity and trending news catering to the Black population.
Figure~\ref{fig:ad_examples} shows two examples of ads and the targeting parameters used by their advertisers in the period when they ran.

Some of the results in Table~\ref{tab:big_table} might appear counter intuitive, for example relatively low skew of Joe Rogan or Roseanne Bar. 
They could be driven by, for example, outdated interest associations, or ~\textit{negative interest}, i.e., one group of users engaging with cross-cutting content in a negative way.
However, regardless of how correct these associations are, our metric describes the \textit{actual} skew of the targeted audience should such targeting criterion be used.

\begin{figure}[t!]
    \includegraphics[width=1\linewidth]{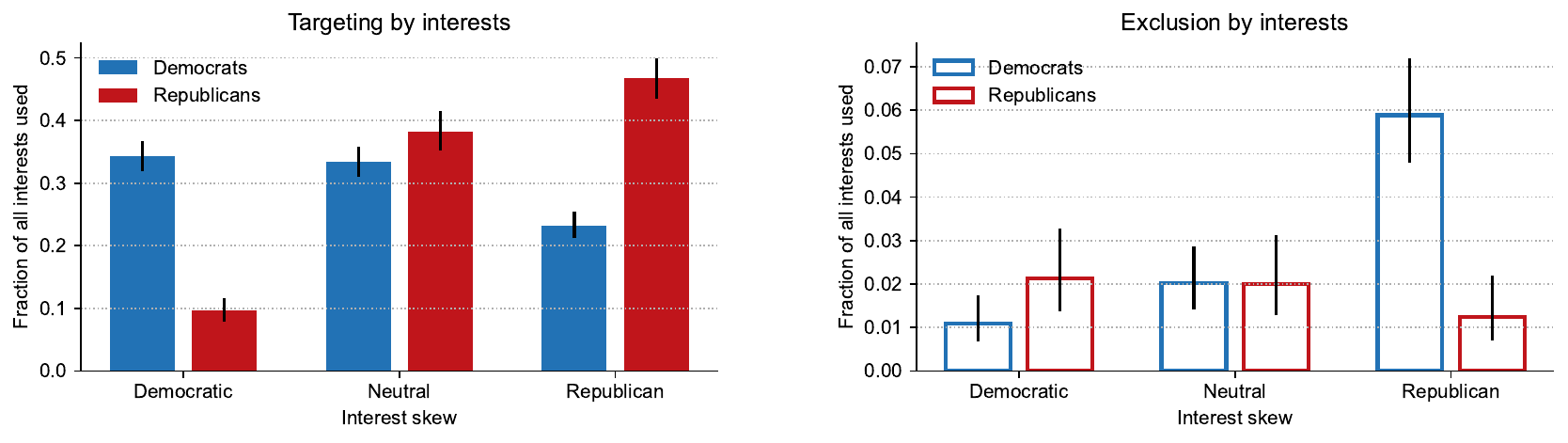}
    \caption{Ad accounts associated with both sides of the spectrum use more congruent than incongruent interests. Republican-leaning advertisers used 4.5x more interests that skew Republican than interests with a Democratic skews. Democrats used 50\% more Democratic skewing interests than Republican skewing.
    Both parties use exclusion more sparingly.}
    \label{fig:interest_skew}
\end{figure}

\subsection{Both Democrats and Republicans use proxies}
We now investigate the prevalence of use of proxies among actors on each side of the political spectrum.
To establish a mapping between an advertising account and its political alignment we rely on data for a subset of advertising accounts reported by WhoTargetsMe (see Section~\ref{sec:whotargetsme}).
Figure~\ref{fig:interest_skew} shows an overview of the results, with interests divided into three equally sized groups (tertiles) along the Republican-Democratic skew:
interests with $S_{RD} < -0.073$ are designated to have a Democratic skew, $-0.73 \leq S_{RD} < 0.063$ are considered Neutral, and those with $R_{RD} \geq 0.063$ -- Republican.

We observe that accounts associated with both sides of the political spectrum tend to target interests that skew in their direction more than those who skew opposite.
This effect is particularly pronounced among Republican-leaning accounts: 46.7\% of interests they use include Facebook users associated with Republican-skewing interests, compared to 9.6\% of interests with a Democratic-skew, nearly a five-fold difference.
Conversely, Democratic-leaning accounts used more Democratic-skewing than Republican skewing interests to target audiences.
We also note that accounts on both sides use fewer interests for exclusion than for inclusion.
Only 9.0\% of interests used by Democrats and 5.4\% used by Republicans were exclusions. 
Nevertheless, Democrats were far more likely to exclude users associated with Republican-skewing interests than those with Democratic skewing interests (5.8\% and 1.1\%, respectively).
That difference for Republican accounts is not statistically significant.

\begin{figure}[t!]
    \includegraphics[width=.58\linewidth]{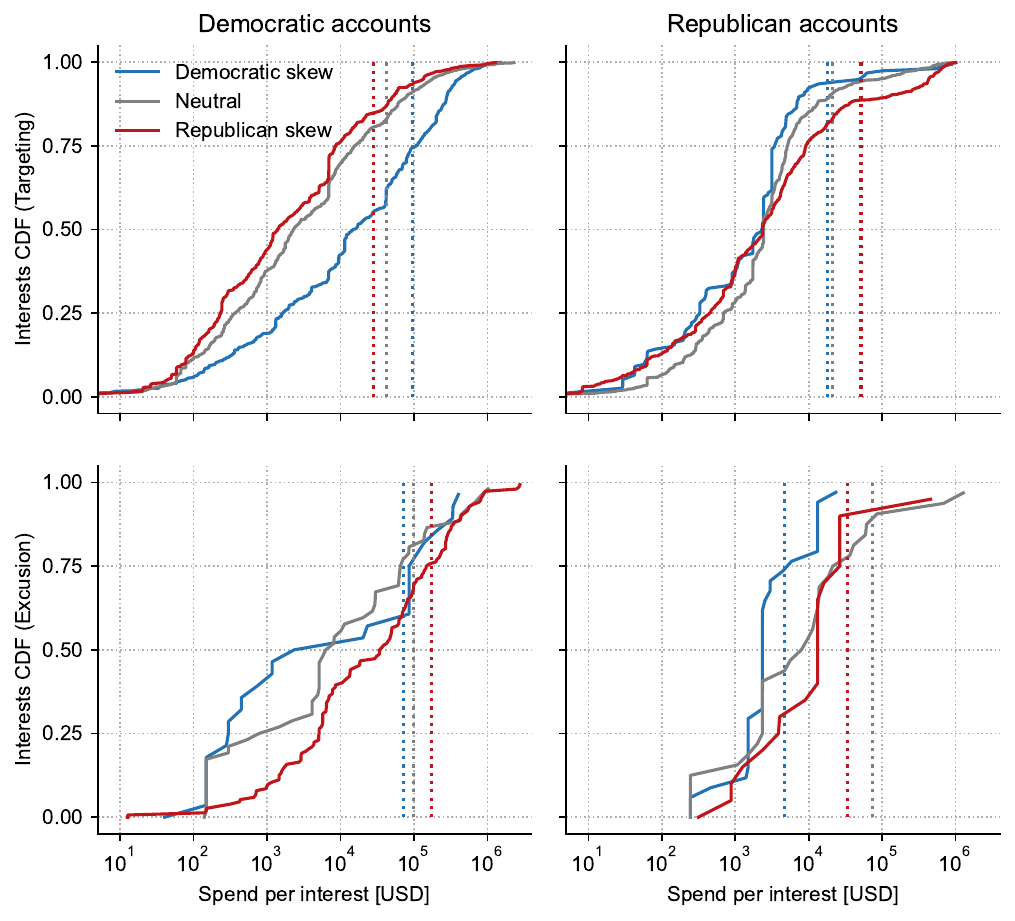}
    \caption{Ad accounts associated with both sides of the spectrum spend on average more targeting interests with aligned skews.}
    \label{fig:spend_skew}
\end{figure}

So far, we have focused on just the count of interest while disregarding how much money the advertisers actually spend on targeting and exclusion using these interests.
We now take a closer look at how advertisers allocated their budgets with respect to different interests.
Figure~\ref{fig:spend_skew} shows the cumulative distribution function (CDF) of spend per interests by Democratic (left panels) and Republican accounts (right panels), both for targeting (top panels) and exclusion (bottom panels).
The dotted lines indicate the mean values.\footnote{A convenient property of a CDF plot allows us to read the median by looking at the value of the x-axis when the plot crosses 0.5 (the mid-point) on the y-axis.}

Beginning with the distribution on targeting spend by Democrats we see that both the median and mean spend per targeted interest is higher for interests with a Democratic skew and the lowest for those with Republican skew.
In fact, Democrat-leaning accounts' median spend on aligned interests is an order of magnitude higher than their spend on the incongruent interests at \$14,696 vs \$1,387, respectively. 
In the bottom left panel, we see that the opposite trends are apparent for exclusion.
Interestingly, both median and mean spend per excluded Republican-leaning interest are higher than the respective values for targeted Democrat-leaning interests. 

Surprisingly, not all of these characteristics hold for the spend distributions among Republican-leaning accounts. 
The median spend per targeted interests regardless of skew is around \$2,300, while the average is still higher for Republican skewing interests, at \$51,524 vs \$17,812 for interests with a Democratic skew.
Perhaps due to limited prevalence of exclusion among Republican-aligned accounts, the distributions in the lower right panel do not paint a clear picture.

\begin{figure}[ht!]
    \includegraphics[width=0.9\linewidth]{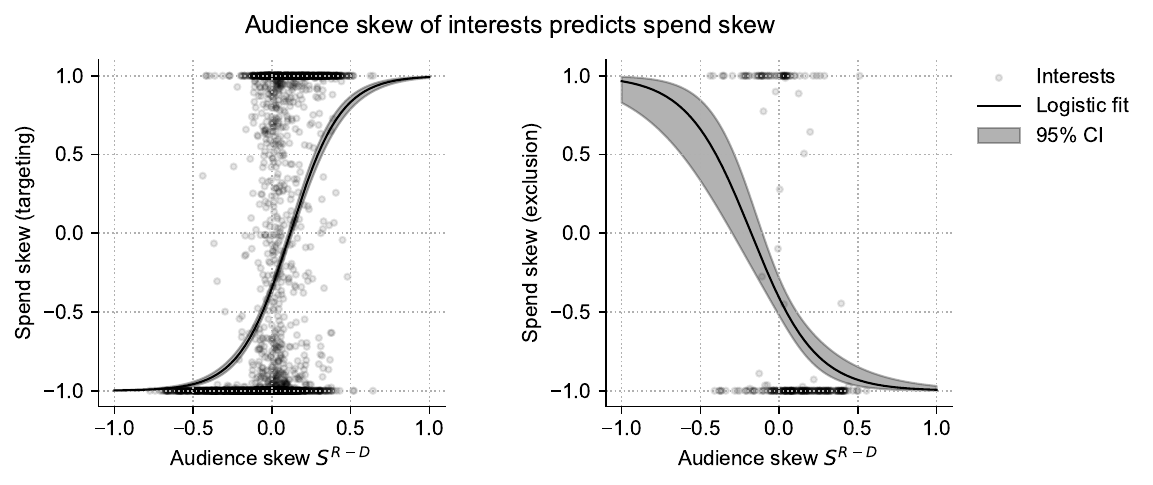}
    \caption{Logistic regression fits show the strong positive relationship between audience skews of targeted interests, and the spend skew between advertisers associated with either party, as well as a negative relationship for excluded interests. This indicates that advertisers on both sides do leverage proxies to reach their supporters and exclude their opponents.}
    \label{fig:reach_spend_skew}
\end{figure}

Still, we note that there is a non-negligible fraction of targeted interests with incongruent skew and excluded aligned interests.
This ``reaching across the aisle'' could be driven by a number of motivations: 
it could indicate a genuine effort at bridging the divide, 
discouraging the other side from voting,
or simply relying on interests that skew in a different direction than anticipated by the advertiser.
Unfortunately, because the Ad Library only reports targeting parameters in weekly aggregate, rather than per-ad, it is difficult to match particular ads to their targeting criteria.
Without that matching it is not always obvious which of these three motivations were the most likely.
We leave the further exploration of reaching across the aisle to future work.

Spending and interest use analysis is finally summarized in Figure~\ref{fig:reach_spend_skew}. 
We introduce a spend skew measure with a formula that resembles the audience skew, i.e., we divide the difference between Republican and Democratic spend on a particular interest by the sum of their spend on that interest.  
Note that this measure is not normalized and that we do not know the affiliation of all advertising accounts. Therefore the value of 0 does not necessarily mean even spend among both parties (but only among the accounts for which we have the affiliation labels). 
We perform a logistic regression fit with audience skew as the independent variable and the spend skew as the dependent variable, to show that the former is predictive of the latter. 
The $R^2$ values for the targeting and exclusion fits are 0.19 and 0.14 and the coefficient associated with the audience skew is significant (see Table~\ref{tab:regression}).
This shows that the audience skew is predictive of the spend skew, and confirms that advertisers on both sides do leverage proxies to reach their
supporters and exclude their opponents.
Nevertheless, the audience skew of interests does not fully explain the targeting strategies of the advertisers.

\begin{table*}[th!]
\centering
\renewcommand{\arraystretch}{1}
\begin{tabular}{@{} p{4cm} *{2}{d{2.5}} @{}}
{\bf Variable} & \mc{\bf Targeting} & \mc{\bf Exclusion} \\
  \midrule
  Intercept & -0.73\sym{***} & -0.87\sym{***} \\
  Audience skew $S^{R-D}$ &  6.25\sym{***} & -4.92\sym{***} \\
    \midrule
    $R^2$ & 0.19 & 0.14 \\
\end{tabular}

\caption{Logistic regression table for predicting spend skew from audience skew. It shows that both parties spend more targeting interests that are congruent in skew (i.e., Democrats spent more targeting interests that reach predominantly Democratic audiences), and excluding interests that are incongruent. $p<0.001^{***}$; $p<0.01^{**}$, $p<0.05^*$.}
\label{tab:regression}
\end{table*}

Taken together, the findings in this section indicate that advertisers use interests-based targeting predominantly---but not exclusively---to reach users who are more likely to be aligned with the advertised messages and, to a lesser extent, exclude those who are not.

\section{Discussion}
We now discuss the ethical considerations taken into account throughout our study, the study's limitations, and conclude with implications of our findings for further regulatory and product changes needed and for future research.

\subsection{Ethics}
Throughout this work, we took careful steps to minimize the impact of our research on all stakeholders.
We limited the strain on Facebook's server infrastructure by enforcing intervals between issuing queries. 
We used real Facebook accounts, one at a time, and we did not attempt to circumvent the rate limits by parallelizing our scraping.
For the collection of targeting information as well as audience insights we relied on public-facing services that are part of the transparency efforts undertaken by Facebook.
They only present identifiable information about advertisers, not individual users.

Previous work found that Facebook did not extract any information from the custom audience files it did not already possess~\cite{venkatadri2019investigating}.
Thus, we do not create additional privacy exposure for individuals whose data is contained within the voter records that we upload to create custom audiences.
Further, we do not actually run any ads, and thus we do not interact with or alter the online experience of the individuals in our custom audiences.
Our analysis did not require collecting any additional non-aggregated user information beyond projected counts. 

We acknowledge that the methods presented in our work could be used by malicious advertisers to find proxy interests according to political skew, race, or ethnicity.
However, our results show that trying to maintain obscurity is futile: mainstream advertisers are already relying on proxies to circumvent Facebook's limits on targeting.
To limit the risk of promoting this practice further, we do not publish the full list of interest skews. 
We believe that by publishing this work we can help encourage Facebook and other ad targeting platforms to take the steps necessary to prevent this kind of abuse.

\subsection{Limitations}
Our work suffers from limitations that stem both from possible data quality issues, as well as simplifying choices we have made during the analysis.
First of all, the targeting information API did not always work reliably, which leads to uneven data quality among advertisers. 
Second, the data is returned in weekly aggregates but collected daily. We normalized spends to account for this discrepancy, but, given the variable delays and selectively missing data, our estimates of spend might not be exact. 
While these two shortcomings might somewhat affect the accuracy of reported spends, we expect that the sorting and direction of skews closely reflects reality.

Further, we relied on manually labeled political affiliations on a selected subset of advertising accounts.
We recognize that the manual labeling process is prone to error and the coverage is far from perfect.
Nevertheless, obtaining higher coverage would only strengthen the results as it would cover more instances of proxy use by partisan actors.
We note that our results do not point to either of the sides of the U.S. political spectrum as a main perpetrator in proxy use. 

We note that our racial (Black, white, and Hispanic) and political labels (Republican, Democrat) are necessarily reductive and do not cover the diverse pool of ethnicities and political stances of U.S. voters.
We consider this work as the first step towards understanding the use of proxies among the political advertisers and the methods we proposed can be extended to measure biases using more labels, and even dimensions beyond race and political views.

Finally, the methods we presented cannot be readily replicated on other platforms.
In our research we rely on using information we obtain through voter records to craft politically or racially homogeneous Custom Audiences and then asking the platform to estimate the prevalence of each interest in a given Custom Audience.
While all major online social media platforms allow advertisers to use tools equivalent to Facebook's Custom Audiences, they do so with caveats.
TikTok and Twitter do not support matching by mailing address, only by phone number, email address, usernames, and advertising IDs.
Only some states publish phone numbers and/or email addresses of their voters, and even then, only a few voters actually provide them.
Google, on the other hand, does support both Custom Audiences based on physical addresses, and intersecting Custom Audiences with inferred interests.
We therefore expect that our method could be applied with minor changes to investigate the skews of targeting interests on Google and we leave the implementation to future work.

\subsection{Implications}
Taken together, our findings demonstrate that the Facebook advertising system continues to allow advertisers to selectively reach or exclude users with certain political views and ethnicities, and that the removal of select interests from the targeting options did not address this problem.
Our findings show that the problem is not just a hypothetical.
We documented that real political advertisers spent millions of dollars in the 2022 midterm elections exploiting proxy targeting to reach---or exclude---select supporters of specific political parties or members of racial and ethnic groups.
The methodology we proposed allows us to directly measure the potential of various targeting criteria to serve as proxies so, as opposed to press reporting on this problem so far~\cite{markay2022americas}, we are not limited to criteria with obvious cultural interpretation.
Although we are not able to measure proxy use in opportunity advertising given the limitations of the Ad Library, our findings of their extensive use in political advertising likely extend to the feasibility of their use in the opportunity advertising context.
Taken together, our work has implications for policy in political and opportunity advertising and platform design, further platform transparency desiderata, and raises a number of questions for future work, including for the platforms themselves.

\subsubsection{Targeted removal of select interests is counter-productive to societal goals}
As discussed in background, public interest entities view the ability to target political messages by racial or ethnic origin, sexual orientation, political opinion, and other inferred personal data as a threat to democratic process with a potential for creating echo chambers,  sowing division and in contradiction with the right of individuals to be informed in an objective, transparent and pluralistic way ~\cite{eu_political_ads, lomas2023ads, eu_political_ads_2024, HRW2022}.
Facebook's removal of ability to target political messages by race, ethnicity and political affiliation, as well as other select topics, and the letter from its employees~\cite{fbemployees2019} is evidence that Facebook also views such dangers as significant.
Furthermore, Facebook's internal documents---as reported on by the Wall Street Journal in 2021~\cite{wsj2021}---explicitly speak of the power of targeting to suppress turnout among certain demographics, and to ``lead to polarization, political disenfranchisement, and information siloing depending on the content, targeting and delivery of the ad”.
Thus, by limiting advertisers' ability to reach politically charged interest-based audiences, Facebook's goal is to reduce the ease with which advertisers could seek to disenfranchise a subset of voters~\cite{wsj2021}.
However, as our work demonstrates, the current approach to such removal may not be sufficiently effective, as advertisers may (and do) circumvent such restrictions through proxy targeting.
Moreover, one may argue that such removal, in fact, has the opposite effect of the one publicly stated in Facebook's goals of increased transparency and user control~\cite{Harbath_Satterfield_2018, Facebook_target_2019pol, Leathern_2020, fb_eu_data_blog}.

First, it concentrates power in political incumbents.
A key argument in 2018 in Facebook's choice not to remove political advertising from its platform was that ``banning political ads on Facebook would tilt the scales in favor of incumbent politicians and candidates with deep pockets. [... ] it would make it harder for people running for local office---who can’t afford larger media buys---to get their message out''~\cite{Harbath_Satterfield_2018}.
Facebook reiterated this argument also in 2022~\cite{Facebook_target_2022}, stating that the targeting is particularly valuable to small advertisers who may not be able to afford broadcast television.
The sentiment for the need of political parties to have equality of opportunity to compete for voters' support is echoed by public interest researchers~\cite{HRW2022}, and even by the Equal Time Rule.\footnote{Section 315 of the Communications Act of 1934 requires that U.S. radio and broadcast TV to provide competing political candidates with equivalent access to present their messages.}
However, what the select removal of targeting criteria, including the easily explainable high-level targeting criteria such as political affiliation, while allowing proxy targeting leads to  the opposite dynamic.
Entities that have larger budgets to acquire data on individuals from data brokers for use in Custom Audiences, those that have more extensive data teams to connect interest data with their proxy power, or those that can otherwise leverage their incumbent positions for proxy targeting efforts, may gain significant political advantage.
Campaign political strategists from both sides of the aisle in the U.S.~were openly discussing their intent to do so~\cite{tolan2022political, corasaniti2021political} in 2021 -- 2022 in response to the removals announced by Facebook coming to force in January 2022.
Furthermore, several recent studies of incumbent data advantage for use in ad targeting in Hungarian election~\cite{HRW2022} and UK election~\cite{uk2018} demonstrate that this concentration of power is not a merely theoretical possibility.

Second, it makes divisive and discriminatory advertising harder to study.
Although the practice of select targeting criteria removal by Facebook in response to concerns began in 2017, our study is the first of its kind to systematically study proxy use without reliance on anecdotal correlations between interests and ethnic or political affinities.
Thus the removal practice, combined with Facebook's current limited Ad Library transparency choices \cite{adlibrarybroken2021}
increases the barrier for effective study of whether racial, ethnic, or other kinds of political manipulation are continuing to take place for public interest researchers independent of Facebook.
Furthermore, the same increase in the difficulty of study applies to the studies of potential discrimination using proxies in the domains of opportunity advertising.

Finally, it makes it more difficult for Facebook users to exercise meaningful control over the personalization of advertising they receive, a desire held by users~\cite{pew2020, gallup_2020, habib2022identifying, hersh2013targeted} and legislators~\cite{lomas2023ads, eu_political_ads, eu_political_ads_2024}, and a stated goal by Facebook~\cite{Leathern_2020}. 
Specifically, the removal of certain interests from available targeting criteria also removes them from ad preference control choices and the user-facing explanations of why they have been shown a particular ad, the two key entry points where users can express their preferences. 
Thus, rather than being able to explicitly specify that the user does not want to receive advertising tailored to their ethnic origin, political affiliation or newspaper readership, whose consequences one may be able to reason about, the user is faced with an insurmountable task of guessing which interests may be  proxies.

Our work thus suggests that the most immediate and feasible solution (albeit with a significant caveat discussed in Section~\ref{sec:ad-delivery-oq}) is the one that European policy makers, civil rights advocacy groups in the U.S. and the E.U., and prominent figures such as Ellen Weintraub and Bill Gates have been advocating for: limiting the targeting capabilities for high-stakes contexts such as political advertising to a very small list of ``safe'' targeting parameters such as broad location and requiring explicit user consent for use of any data beyond that~\cite{gates2019}.
Another approach would be to limit the available targeting criteria to a very short list of those directly relevant to political engagement and give users robust controls associated with those.

\subsubsection{Balancing fine-grained targeting and societal desiderata}
As a middle ground to significantly limiting the targeted advertising in political and opportunity domains, platforms can aim to develop new approaches that balance their desire to provide political advertisers, including small ones, with enhanced abilities to engage audiences in the political process, with effective transparency efforts enabling independent oversight of the political advertising for public interest entities, and meaningful controls for individual users.

\parc{Platform work on proxy identification and removal.}
A path for platforms to keep the fine-grained targeting capabilities at the variety level of today while satisfying the societal desiderata is for the platforms to take it upon themselves to continuously identify and remove proxy attributes and attribute combinations among their offered targeting capabilities.
These decisions should not be limited to proxy attributes whose names clearly indicate skew, and should be based on more systematic methods, such as those we presented in this work. 
However, for this approach to convincingly meet the societal and regulatory goals, and to gain public trust, platforms' decisions and data on which they are based should be available for independent third party and public interest scrutiny.
This is not the case today: despite thousands of attribute removals over the course of seven years, Facebook did not make the targeting attributes removed, or the attributes left, nor the specific reasoning for their choices, available~\cite{nieman2022}.

\parc{Transparency of targeting.}
Making the targeting choices of political and opportunity advertisers available to public interest scrutiny has long been a key request of civil society organizations from the platform~\cite{wapost2019, CLUFE2021}.
Facebook itself makes the claim that ``information about advertisers’ targeting choices is critical to understanding the impact of digital advertising on elections and social discourse''~\cite{king2022targeting}.

However, currently, Facebook's Ad Library presents the
targeting criteria used by political advertisers as a weekly aggregate across all of that advertiser's campaigns, rather than per advertisement or ad campaign.\footnote{Furthermore, there is no official maintained or documented API to access the targeting information, an omission that limits inquiry and should be remedied.}
For opportunity advertisers, the targeting criteria are not part of the Ad Library at all.
Facebook cites user privacy concerns as the key reason behind withholding that data~\cite{facebook2021privacy, king2022targeting}.
For example, if targeting criteria for each ad are publicly available, then if an attacker knows that an ad was targeted using criterion $X$, and can also observe that user $u$ has interacted with this ad, then the attacker can infer that user $u$ matches criteria $X$~\cite{korolova2011privacy}. 
Such an attack becomes particularly worrisome given the availability and use of proxy interests that could thus imply sensitive information about the user.

One path to approaching the transparency-privacy trade-off is through deploying legal protections to minimize risk of abuse from those granted access. 
This is an approach that Facebook began to pursue in May 2022 through its Facebook Open Research and Transparency (FORT) platform~\cite{fb_fort_data}. Specifically, ``eligible university-affiliated researchers'' can apply to FORT to access targeting data at the ad level, if Facebook approves their study's research scope and objectives~\cite{fb_fort_terms}, and if the researchers accept the terms in its Research Platform Addendum~\cite{fb_fort_addendum},\footnote{We did not seek access to this data as we find that the terms of the Addendum are incompatible with academic freedom.} which include provisions prohibiting privacy violations.
Although a viable path, it has the shortcoming of leaving the control of what research can be done, and by whom, in the platform's hands, which may stifle the ability to perform and publish findings with unfavorable outcomes to the platform.

Another potential path to mitigate the privacy issues is by changing the visibility of user interactions with paid content, particularly paid political content.
For example, rather than making the user interactions with political ads visible to the entire world (as is done currently), the default privacy setting for such interactions could be set to private or friends-only. The restriction would make the interactions much harder to access at scale for potential attackers; and thus make it difficult and not cost effective to violate privacy using the transparency tools.
A complementary approach, would be to make clear to the users affirmatively opting-in to public interactions what information about them may be revealed.
Developing ways to make the targeting explanations accurate, comprehensive and interpretable by the users, and ways to provide them in context~\cite{im2023less}, is an important area for future work.

\parc{Beyond targeting transparency.}
We take a further step back to re-imagine how to expand the Ad Library to make it an effective tool for research on the questions that inspired its creation: (1) do advertisers exploit the targeting options to reach relatively narrow segments of users that disproportionately share a characteristic that can be negatively exploited in the political process or in opportunity advertising, and (2) do advertisers send conflicting messages to separate groups of users? 
We suggest that rather than approaching it as a question of merely making some data available and placing the burden of its analysis on the (usually) less technically sophisticated and resourced public interest researchers, platforms can play an active and constructive role in enabling such research.
For example, platforms could develop and deploy privacy-preserving ways to examine skew in the composition of an advertiser's targeted audience along a variety of characteristics (and their intersections), and highlight such advertisers along with the characteristics for public interest scrutiny.
Methods for doing so already exist in other domains~\cite{rogers2020linkedin, alao2021meta}.
Furthermore, platforms could develop techniques for flagging when ad message tailoring techniques cross from the valuable to the potentially harmful and manipulative, and bring those messages to public interest scrutiny.
The holistic view that the platform has of the messages used as well as their evolution over time can make this task feasible.
Finally, the platforms, while leveraging user reports~\cite{lam2023sociotechnical}, could develop new techniques for identifying when a combination of an ad's message with the targeting chosen is manipulative or pernicious, even when each is seemingly benign on its own.
Platforms have hinted at efforts to do so, but have not made the techniques nor the findings public~\cite{Schrage2017}.
These approaches can be complementary to the others, can rely on privacy and verification techniques that offer better trade-offs between privacy and utility for data analyses rather than data releases, and can lead to a constructive dialogue between platforms, users and civil society that expands the set of possible solutions to the benefit of all stakeholders.

\parc{Center the needs of end users.}
Finally, platforms should continue improving transparency in ways that center the needs of their end-users.
In particular, individuals should have meaningful overview and control of the information that is used to personalize delivery of political (and opportunity) advertising to them and be protected from manipulation. 
According to PEW research in 2020, as many as 77\% of Americans said that the practice of using their online activities to show them ads from political campaigns is not very or not at all acceptable~\cite{pew2020}.
The statistics are echoed by a Gallup study that indicates the majority of Americans do not want to be micro-targeted for political ads~\cite{gallup_2020}.
In the EU, this desiderata in the political ads context has recently been mandated by European regulation~\cite{eu_political_ads_2024} but the pathway for practically achieving it remains an open question.

Initial survey based methods have begun to explore user needs related to advertising controls broadly~\cite{habib2022identifying}, and platforms should expand and tailor such studies to political and opportunity advertising.
In particular, in addition to the reasons previously discussed related to discrimination in opportunity advertising,
Facebook frames their choices for targeting attribute removal as those that relate to ``topics people perceive as sensitive''~\cite{Facebook_target_2022}. 
However, what is, in fact, sensitive for each individual, may be individual- and context-dependent~\cite{lee2023and}.
Furthermore, for individuals, reasoning about sensitivity may be easier at the high-level category levels, rather than at the level of categories whose proxy power, individually and in combination, the user does not know.
Third-party organizations, such as WhoTargetsMe, have begun to re-package the information available through the Ad Library and combine it with ad targeting explanations collected through browser plugins to make such understanding possible~\cite{whotargetsme_extension}. 
However, the onus of these kinds of transparency and controls should be on the much better-resourced platform, and innovation in this space remains an urgent direction for future work.

\subsubsection{Role of algorithmic ad delivery when targeting is limited}\label{sec:ad-delivery-oq}
A consequence of removing fine-grained and other targeting abilities of advertisers is that it shifts the power over how political speech and opportunity advertising is distributed from the advertiser to the platform.
Facebook in particular encourages greater reliance on its algorithmic tools, such as Lookalike Audiences, MetaAdvantage+ audience when removing the targeting abilities~\cite{Facebook_target_2020, Facebook_target_2022b, Facebook_target_2024, Chouaki_Bouzenia_Goga_Roussillon_2022}.
As we discussed in Section~\ref{sec:ad_delivery}, algorithmic identification of the right audience and machine learning driven ad delivery optimization, are processes that can (and do) lead to echo chambers in political ad delivery~\cite{ali2021ad, bar2024} and discrimination in opportunity advertising~\cite{ali2019discrimination, imana2021auditing, kaplan2022measurement, sapiezynski2022algorithms, imana2024auditing}, even when the advertiser targets a politically or racially balanced audience.
Furthermore, recent work shows that for ads with greater reliance on algorithmic targeting tools, the user-facing explanations and controls are inaccurate and ineffective~\cite{castleman2024why}.
Thus, the question of reconciling the tension between granularity of targeting available to advertisers and ceding the control of targeting and delivery to the platform, and, hence, the transparency desiderata for the implementations of the ad delivery algorithms in political and opportunity advertising domains is an important question for future work.

\section{Acknowledgements}
This work work was funded in part by National Science Foundation grants 
CNS-1956435, 
CNS-2344925, 
CNS-1955227, 
and 
CNS-2318290. 
We thank the CSCW reviewers for their thoughtful and detailed comments that have helped improve this work.

\bibliographystyle{ACM-Reference-Format}
\bibliography{bibliography}

\appendix


\balance
\end{document}